\documentclass[sigconf]{acmart}
\usepackage{multirow}
\usepackage{tabularx}
\usepackage[nolist,nohyperlinks]{acronym}
\usepackage{lipsum}  
\usepackage{color, colortbl}
\usepackage{kg-macros}
\usepackage{makecell}

\usepackage{subfig}

\definecolor{gry}{gray}{0.9}
\definecolor{LightGreen}{rgb}{0.9,1,0.8}

\newcommand{\stab}{\vspace{1.2ex}\noindent}
\newcommand{\stitle}[1]{\stab\noindent{\bf #1}}
\definecolor{LightCyan}{rgb}{0.88,1,1}

\newcolumntype{a}{>{\columncolor{gry}}c}
\newcolumntype{b}{>{\columncolor{white}}c}
\newcommand{\most}[1]{\colorbox{green!20}{#1}}
\newcommand{\least}[1]{\colorbox{red!20}{#1}}

\usepackage{arydshln}

\AtBeginMaketitle{%
\def\EDBTISSN{2367-2005}
\def\EDBTISBN{978-3-98318-104-9}
\setcopyright{none}
\copyrightyear{\copyright\ 2026 Copyright held by the owner/author(s).
  Published on \url{OpenProceedings.org} under ISBN \EDBTISBN, series ISSN \EDBTISSN.
  Distribution of this paper is permitted under the terms of the
  Creative Commons license CC-by-nc-nd 4.0}
\acmYear{2026}
\acmDOI{}
\acmISBN{}

\acmConference[EDBT '26]{Extending Database Technology}{24-27 March 2026}{Tampere (Finland)}
\settopmatter{printacmref=false, printccs=false, printfolios=false}

\newsavebox{\ximagebox}
\newlength{\ximageheight}
\newsavebox{\xglyphbox}
\newlength{\xglyphheight}
\newcommand{\xbox}[1]%
  {\savebox{\ximagebox}{#1}%
  \settoheight{\ximageheight}{\usebox{\ximagebox}}%
  \savebox{\xglyphbox}{\color{white}\char32}%
  \settoheight{\xglyphheight}{\usebox{\xglyphbox}}%
  \raisebox{\ximageheight}[0pt][0pt]{\raisebox{-\xglyphheight}[0pt][0pt]{%
    \makebox[0pt][l]{\usebox{\xglyphbox}}}}%
    \usebox{\ximagebox}%
    \raisebox{0pt}[0pt][0pt]{\makebox[0pt][r]{\usebox{\xglyphbox}}}}

\newsavebox{\LogoBox}
\sbox{\LogoBox}{\includegraphics[height=1cm]{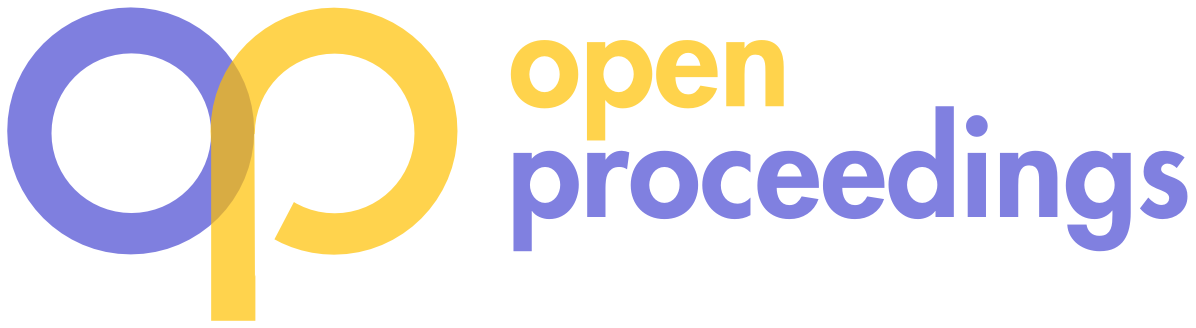}}

\fancypagestyle{firstpagestyle}{%
  \fancyhf{}%
  \fancyfoot{}%
  \fancyhead[L,C]{}
  \fancyhead[R]{\href{https://OpenProceedings.org/}{\raisebox{15pt}[0pt][0pt]{\xbox{\usebox{\LogoBox}}}}}
  }

\hypersetup{%
  pdfcopyright={CC-by-nc-nd},
  pdfpublisher={OpenProceedings.org -- University of Konstanz, Germany},
  pdfcreator={LaTeX with acmart, hyperref, and EDBT modifications},
  pdfvolumenum={29},
  pdfissuenum={2},
  pdfisbn={\EDBTISBN},
  pdfissn={\EDBTISSN}
}

}% \AtBeginMaketitle

\begin{document}
\begin{acronym}
    \acro{RAG}{Retrieval-Augmented Generation}
    \acro{KG}{Knowledge Graph}
    \acro{CSV}{Comma Separated Values}
    \acro{NLI}{Natural Language Inference}
    \acro{NLP}{Natural Language Processing}
    \acro{LLMs}{Large Language Models}
    \acro{LLM}{Large Language Model}
    \acro{QA}{Question Answering}
    \acro{RAG}{Retrieval-Augmented Generation}
    \acro{ML}{Machine Learning}
    \acro{IR}{Information Retrieval}
    \acro{HTML}{HyperText Markup Language}
    \acro{RoPE}{rotary position embeddings}
    \acro{MTEB}{Massive Text Embedding Benchmark}
    \acro{LoRA}{Low-Rank Adaptation}
    \acro{MRL}{Matryoshka Representation Learning}
    \acro{RLHF}{Reinforcement Learning with Human Feedback}
    \acro{SFT}{Supervised Fine-Tuning}
    \acro{GQA}{Grouped-Query Attention}
    \acro{SWA}{Sliding Window Attention}
    \acro{RS}{Rejection Sampling}
    \acro{DPO}{Direct Preference Optimization}
    \acro{AI}{Artificial Intelligence}
    \acro{KB}{Knowledge Base}
    \acro{KG}{Knowledge Graph}
    \acro{KGs}{Knowledge Graphs}
    \acro{KBs}{Knowledge Bases}
    \acro{HTML}{Hypertext Markup Language}
    \acro{MLM}{Masked Language Modeling}
    \acro{SERP}{Search Engine Results Page}
    \acro{QA}{Question Answering}
    \acro{DKA}{Direct Knowledge Assessment}
    \acro{GIV}{Guided Iterative Verification}
    \acro{RQ}{research question}
    \acro{IQR}{Inter Quartile Range}
\end{acronym}

\title{Benchmarking Large Language Models for Knowledge Graph Validation}

\author{Farzad Shami}
\email{farzad.shami@aalto.fi}
\orcid{0009-0004-8174-0082}
\affiliation{%
  \institution{Aalto University}
  \city{Espoo}
  \country{Finland}
}

\author{Stefano Marchesin}
\email{stefano.marchesin@unipd.it}
\orcid{0000-0003-0362-5893}
\affiliation{%
  \institution{University of Padua}
  \city{Padua}
  \country{Italy}
}

\author{Gianmaria Silvello}
\email{gianmaria.silvello@unipd.it}
\orcid{0000-0003-4970-4554}
\affiliation{%
  \institution{University of Padua}
  \city{Padua}
  \country{Italy}
}

\renewcommand{\shortauthors}{Shami et al.}

\begin{abstract}
Knowledge Graphs (KGs) store structured factual knowledge by linking entities through relationships, crucial for many applications. These applications depend on the KG's factual accuracy, so verifying facts is essential, yet challenging. Expert manual verification is ideal but impractical on a large scale. Automated methods show promise but are not ready for real-world KGs. Large Language Models (LLMs) offer potential with their semantic understanding and knowledge access, yet their suitability and effectiveness for KG fact validation remain largely unexplored.

In this paper, we introduce \approach, a benchmark designed to evaluate LLMs for KG fact validation across three key dimensions: (1) LLMs internal knowledge; (2) external evidence via Retrieval-Augmented Generation (RAG); and (3) aggregated knowledge employing a multi-model consensus strategy.
We evaluated open-source and commercial LLMs on three diverse real-world KGs.
\approach also includes a RAG dataset with 2+ million documents tailored for KG fact validation. %Additionally, we offer an interactive exploration platform for analyzing verification decisions.

The experimental analyses demonstrate that while LLMs yield promising results, they are still not sufficiently stable and reliable to be used in real-world KG validation scenarios. Integrating external evidence through RAG methods yields fluctuating performance, providing inconsistent improvements over more streamlined approaches -- at higher computational costs. Similarly, strategies based on multi-model consensus do not consistently outperform individual models, underscoring the lack of a \emph{one-fits-all} solution. These findings further emphasize the need for a benchmark like \approach to systematically evaluate and drive progress on this difficult yet crucial task.
\end{abstract}


\maketitle
\begingroup

\section{Introduction}
\acp{KG} are machine-interpretable, directed, labeled multigraphs in which nodes represent entities or concepts, and edges denote typed semantic relations. They provide a structured representation of real-world knowledge, enabling reasoning, integration, and querying across information sources~\cite{Knowledge_graphs,10.1145/3447772,jiang2023evolutionknowledgegraphssurvey}. \acp{KG} have been deployed in a wide
range of applications~\cite{10.1145/3447772, ieeesurveykg}, including: (1) web search for semantic understanding of queries and content~\cite{10.1145/3329781.3332266,10.1145/775152.775250}; (2) e-commerce, for recommendation~\cite{sharma2018scaling} and conversational agents~\cite{pittman2017conversational}; (3) social networks, for modeling user interests~\cite{he2016building, 10.1145/3329781.3332266}; and (4) other domains such as finance~\cite{8731350}, transport~\cite{henson2019using}, and energy~\cite{kg-carbon}.
However, the effectiveness of downstream applications depends on the accuracy of the \ac{KG}'s facts. Each individual piece of knowledge, which is typically represented as an \texttt{<S,P,O>} triple (i.e., \gedge{Subject}{Predicate}{Object}), must be factually correct. In addition, the reliability of the entire \ac{KG} depends not only on the correctness of these atomic facts, but also on the way they are interconnected~\cite{deshpande_etal-2013,pujara_etal-2017}.\footnote{We use the terms fact, statement, and triple interchangeably depending on the context.}

%\begin{figure*}[t!]
%    \centering
%    \includegraphics[width=\linewidth]{figures/KG_Validation}
%    \caption{An example of the \approach \ac{KG} Validation Pipeline.}
%    \Description{\ac{KG} Validation Pipeline.}
%    \label{fig:KG-validation}
%\end{figure*}

A crucial step after the creation of the \ac{KG} is assessing the veracity of its facts~\cite{10.1145/3447772,10.1145/3749838}. This involves determining how accurately the data reflects real-world entities, relationships, and phenomena. Fact validation presents a significant challenge and is expensive~\cite{marchesin_silvello-2024,marchesin_silvello-2025}. The most reliable option involves manual or computer-assisted annotation by human experts~\cite{10.1007/978-3-030-64452-9_35, WYSOCKA2024104724}. However, this process is extremely time-consuming~\cite{gao_etal-2019,ojha-talukdar-2017-kgeval}. Since experts often need to audit facts relying on multiple external reference sources, in large-scale \acp{KG} (e.g., DBpedia~\cite{dbpedia-large-scale} or YAGO~\cite{10.1145/1963192.1963296}), verifying each individual triple can take several minutes, making manual inspection and correction of errors infeasible at scale. 

As a result, automated fact-checking methods~\cite{10.1145/3269206.3269308,GERBER201585,8215568,syed2019copaaldemo,10.1145/2488388.2488425}, often based on rules and enforceable constraints~\cite{10.1145/2488388.2488473,10.1007/978-3-319-07443-6_29}, have emerged as more scalable alternatives to address the time and cost limitations of human-based solutions. While these methods are effective for well-defined and frequently occurring facts \cite{kim-choi-2020-unsupervised}), they fall short when it comes to generalizing across the wide variety of facts found in real-world KGs. %Indeed, the core challenge is to automatically generate rules and constraints that are broadly applicable and is largely unresolved. 
Manual definition, on the other hand, is both difficult and expensive. Therefore, (semi-) automatic methods that extract rules and constraints can be employed. Nonetheless, these methods predominantly cater to rules that identify frequent positive instances and encounter difficulties with cases pertaining to infrequent facts or necessitate the application of negative rules~\cite{8509329}.

These limitations have led to the adoption of fact-checking systems with machine/deep learning solutions~\cite{10.1162/tacl_a_00454}. In this realm, a viable approach could be to utilize \acp{LLM} for fact-checking, as they have demonstrated near-human-level performance on various tasks~\cite{pan_et_al:TGDK.1.1.2}. Within this framework, \acp{LLM} offer various advantages: they can extract contextual information from text, comprehend the semantics of statements, and possess an extensive internal knowledge base~\cite{petroni, wang}. However, current \acp{LLM} generate hallucinated and unfaithful responses~\cite{DBLP:conf/naacl/SunXZLD24}. Additionally, recent work has highlighted that \acp{LLM} are particularly problematic for fact validation tasks, exhibiting systematic biases and knowledge gaps that can affect their reliability~\cite{sun-etal-2024-head}. To combat the limitations caused by knowledge cutoff and hallucination in \acp{LLM}, current systems built on top of \acp{LLM} often implement a \ac{RAG} approach in which the \ac{LLM} is supplemented with data from external sources to improve their responses~\cite{khaliq-etal-2024-ragar}.
However, despite all the recent progress in \ac{LLM} research and the capability of \acp{LLM} to tackle a wide range of tasks, there appear to be no existing benchmarks specifically measuring the performance of \acp{LLM} in \ac{KG} fact validation \cite{10.1145/3749838}. 

Hence, we present \approach, a general-purpose benchmark designed to assess LLMs in the validation of \ac{KG} facts across three principal dimensions: (1) LLM internal knowledge; (2) external evidence through Retrieval-Augmented Generation (RAG); and, (3) synthesized knowledge from multiple models. %, akin to the ``wisdom of the crowd'' strategy. 

\approach relies on a validation pipeline that transforms structured triples into natural language statements for evaluation of their factual accuracy. %Figure~\ref{fig:KG-validation} illustrates the core steps of the validation procedure: beginning with \ac{KG} entities and relations, it derives structured facts in a triple format, checks them against reliable sources, and ultimately calculates accuracy scores.
\textcolor{black}{The validation procedure begins with \ac{KG} entities and relations, derives structured triples, checks them against reliable sources, and calculates accuracy scores.}

\approach is driven by the following \acp{RQ}:

\begin{itemize}[noitemsep,topsep=0pt,parsep=0pt,partopsep=0pt,leftmargin=18pt]
\item[\textbf{RQ1:}] How effective are \acp{LLM} at \ac{KG} fact-checking when relying only on their internal knowledge?
\item[\textbf{RQ2:}] Does external evidence improves the ability of \acp{LLM} to fact-check \acp{KG}?
\item[\textbf{RQ3:}] Does aggregating predictions from multiple \acp{LLM} lead to more reliable validation of KG facts?
\end{itemize}

RQ1 targets a recent debate concerning \acp{LLM} functioning as \acp{KB}, aiming to evaluate how factual and complete the internal knowledge of an \ac{LLM} is, for both previously seen and unseen knowledge~\cite{he-etal-2025-language, abs-2402-14273, zheng2024reliablellmsknowledgebases}. We do not prompt the LLM to retrieve knowledge to evaluate its completeness and accuracy. Instead, we ask the LLM to judge the accuracy of externally provided facts, which requires it to depend solely on its internal knowledge. Our focus is directed towards this research approach, acknowledging that studies indicate querying an \ac{LLM} for the verification of information accuracy produces more favorable outcomes compared to prompting it to generate or assess its own content~\cite{kamoi-etal-2024-llms,kumar2024traininglanguagemodelsselfcorrect}.

RQ2 targets the effectiveness of augmenting \acp{LLM} with external evidence to improve KG fact-checking, contributing to ongoing discussions around \ac{RAG} and its role in factual verification~\cite{khaliq-etal-2024-ragar, yue-etal-2024-retrieval, russo2024facefactsevaluatingragbased}. While classical \ac{RAG} approaches often outperform \acp{LLM} that rely solely on internal knowledge, recent findings indicate that RAG effectiveness can diminish in complex or multi-turn settings, where context management and evidence selection become more error-prone~\cite{laban2025llmslostmultiturnconversation}. Moreover, integrating external evidence can introduce contextual bias, where the model overly trusts retrieved content~\cite{leng2024longcontextragperformance}. %regardless of its factual quality
With \approach, we aim to foster research on whether and under what conditions external evidence helps \ac{KG} fact validation, to what extent, and under which conditions. 

RQ3 targets a growing body of work investigating whether aggregating outputs from multiple \acp{LLM} can lead to more accurate or reliable factual verification~\cite{sciadv.adp1528,chen2025harnessingmultiplelargelanguage}. While individual LLMs may vary in factual accuracy, reasoning patterns, and susceptibility to hallucinations, recent studies suggest that combining multiple models -- via voting, consensus, or arbitration mechanisms -- can mitigate individual model biases and increase robustness~\cite{xue-etal-2023-dynamic, wan-etal-2025-reasoning}. However, this approach introduces its own challenges, including disagreement resolution, scaling cost, and the risk of amplifying shared misconceptions among models trained on overlapping data. \approach can help explore whether ensemble-style reasoning from multiple \acp{LLM} can improve the reliability of KG fact-checking. % especially under scenarios where models partially disagree or bring complementary knowledge to the task.
\paragraph{\textbf{Contributions}}
% PREVIOUS VERSION:
% We propose \approach, a benchmark designed for \ac{KG} fact verification using \acp{LLM}, which comes with several advantages: 
We propose \approach, a benchmark for \ac{KG} fact validation using \acp{LLM}, which comes with several advantages:
\begin{itemize}[leftmargin=*]
\item[(1)]
\approach integrates various \acp{LLM} for KG fact validation. The benchmark evaluates these models using both their internal knowledge and external evidence through \ac{RAG}. It also explores consensus-based verification via majority voting strategies. Experiments with mid-sized (7–9B parameters) and commercial LLMs highlight the challenges of the task.
\item[(2)]
\approach is built upon three real-world KG datasets: \textit{FactBench}~\cite{GERBER201585}, \textit{YAGO}~\cite{ojha-talukdar-2017-kgeval}, and \textit{DBpedia}~\cite{Marchesin_Silvello_Alonso_2024}, covering broad spectrum of knowledge, ranging from everyday facts to complex, domain-specific information, ensuring a diverse and representative evaluation of fact validation capabilities.

\item[(3)]
\approach includes a large-scale \ac{RAG} dataset featuring several questions paired with corresponding Google \acp{SERP}. The dataset comprises 2M+ documents covering a broad range of factual information, making it one of the most comprehensive and publicly available RAG resources for KG fact validation. \textcolor{black}{\approach includes a mock API that simulates real search APIs, allowing users to reproduce data retrieval, test retrieval methods, and extend RAG methods without direct access to search engines.}

\item[(4)]
A dedicated web application (\url{https://factcheck.dei.unipd.it/}), enabling users to visually explore and analyze each step of the verification process, also featuring error analysis modules that categorize reasoning errors, enabling systematic identification of LLM limitations in fact-checking scenarios. %The platform supports researchers and practitioners in examining how different LLMs interpret evidence, compare reasoning strategies, and understand verification outcomes. It also features error analysis modules that categorize reasoning errors, enabling systematic identification of LLM limitations in fact-checking scenarios.

\item[(5)]
\approach enables comprehensive evaluation by combining performance metrics with resource usage analysis. Model predictions are evaluated against gold-standard labels to assess accuracy and reliability. The benchmark also tracks computational costs (inference time and token usage). % -- for each LLM, offering detailed insights into their efficiency and scalability in KG fact-checking tasks.
\end{itemize}

Evaluation with different methodologies and datasets highlights the difficulty and inherent complexity of the fact validation task in \ac{KG}.
The main insights of our work are three-fold:
First, while \acp{LLM} show promising capabilities in \ac{KG} fact validation, they are still far from being reliably deployed in real-world validation scenarios.
Second, integrating external knowledge through \ac{RAG} yields fluctuating performance, providing inconsistent improvements over more streamlined approaches at significantly higher computational costs.
Finally, consensus-based strategies using multiple models are unable to consistently outperform individual models. %This shows that there is no universal solution and highlights the need for careful benchmarking to make progress on this important challenge.
Altogether, these results highlight the task’s difficulty and complexity, underscoring the need for a dedicated benchmark to drive progress.

\paragraph{\textbf{Outline}}
The rest of the paper is organized as follows. 
In Section~\ref{sec:related_work}, we review related work on automated \ac{KG} fact-checking and benchmark development. %, and evaluation metrics. 
In Section~\ref{sec:approach}, we introduce the \approach benchmark. % and detail our methodologies using both \ac{LLM}'s internal knowledge and external evidence sources. 
We detail the \approach construction in Section~\ref{sec:construction}, covering both dataset selection and \ac{RAG} corpus creation. Section~\ref{sec:setup} outlines the experimental setup, with results discussed in Section~\ref{sec:performance_report}. \textcolor{black}{Section~\ref{sec:qualitative_error_analysis} provides a qualitative error analysis of failure cases.}
Finally, in Section~\ref{sec:remarks}, we draw final remarks.

\begin{table*}[t!]
\footnotesize
  \caption{\textcolor{black}{Comparative analysis of Internal KG-Based versus External Evidence-Based fact-checking mechanisms.}}
  \label{tab:internal_vs_external}
  \resizebox{\textwidth}{!}{%
  \begin{tabularx}{\textwidth}{@{} l X X @{}}
    \toprule
    \textbf{Feature} & \textbf{Internal KG-Based Fact Checking} & \textbf{External Evidence-Based Fact Checking} \\
    \midrule
    % \textbf{Core Principle} 
    %     & \textit{Coherence}: A fact is likely true if it is structurally consistent with the existing patterns in the graph. 
    %     & \textit{Correspondence}: A fact is true if it aligns with authoritative information found in external world sources. \\
    \textbf{Principle} & \textit{Coherence}: Consistent with graph patterns. & \textit{Correspondence}: Aligns with external sources. \\

    \textbf{Primary Evidence} 
        & Graph topology, paths, and flow networks 
        % (e.g., KStream). 
        & Unstructured text, webpages, and search snippets \\
        % (e.g., DeFacto). \\
    \textbf{Assumption} & 
        %\textit{Local Closed World Assumption}: 
        % Assumes local completeness to derive negative signals from missing links.  
        Derives negative signals from missing links based on local completeness.&
        Missing links are verified against external data under incompleteness.\\
    
    \textbf{Mechanism} & Path mining, link prediction. & IR, NLP, RAG. \\

    \textbf{Handling Negatives} & Synthesized via sampling strategies (e.g., \cite{kim-choi-2020-unsupervised}). 
    & Retrieval failure or contradiction. \\
    % \textbf{Pros \& Cons} & 
    %     \textbf{(+)} High computational efficiency, structural consistency. 
    %     \newline 
    %     \textbf{(-)} Cannot detect systematic graph errors; limited by sparsity. & 
        
    %     \textbf{(+)} Validates against real-world data; handles novel facts. 
    %     \newline 
    %     \textbf{(-)} High latency; dependent on retrieval quality and source reliability. \\
    \textbf{Trade-offs} & \textbf{(+)} Fast, Consistent. \textbf{(-)} Misses graph errors. & \textbf{(+)} High validity. \textbf{(-)} Slow, source-dependent. \\

    \textbf{Examples} & 
        KStream~\cite{8215568}, PredPath~\cite{SHI2016123}, COPPAL~\cite{10.1007/978-3-030-30793-6_36}. &
        DeFacto~\cite{GERBER201585}, KGValidator~\cite{boylan2024kgvalidatorframeworkautomaticvalidation}, \approach (Ours). \\
    \bottomrule
  \end{tabularx}
  }
\end{table*}

\section{Related Work}\label{sec:related_work}
\subsection{Automated KG Fact Checking}
Fact-checking methods can be categorized into approaches that directly utilize the \ac{KG} to find a supporting path for the given statements~\cite{8215568, 10.1007/978-3-030-30793-6_36, kim-choi-2020-unsupervised,SHI2016123} and others relying on external reference sources to find supporting or conflicting evidence~\cite{GERBER201585, 10.1145/3269206.3269308}. \textcolor{black}{Table~\ref{tab:internal_vs_external} represent comparative analysis of these two paradigms.}

\paragraph{\textbf{(1) Internal KG-Based Fact Checking}}
Knowledge Stream (KStream) and Relational Knowledge Linker (KLinker)~\cite{8215568} are unsupervised, network-flow-based approaches designed to assess the truthfulness of factual statements expressed as \texttt{<S,P,O>} triples. KStream models a \ac{KG} as a flow network, where the path carries flow from a subject to an object to support or refute a given statement. KLinker, on the other hand, focuses on discovering relational paths that link entities to each other. COPPAL~\cite{10.1007/978-3-030-30793-6_36} proposes a corroborative meta-path to find statement-supporting paths. These approaches focus only on positive evidential paths and are heavily restricted due to the incomplete nature of KGs. Approaches like PredPath~\cite{SHI2016123} attempt to utilize both negative and positive paths to cover a broader range of factual statements. PredPath assigns weights to discriminative predicate paths by considering only correct examples, ignoring counterexamples. This can lead to improperly weighted rules. 
In addition, \citet{kim-choi-2020-unsupervised} presents an unsupervised rule-based approach that significantly outperforms the state-of-the-art unsupervised approaches in this area. They calculate a truth score for the given statement by finding positive and negative evidential paths in a \ac{KG}, generating examples for the training phase, creating a model for learning from positive and negative rules, and scoring the triple based on established evidence.

While these methods are effective, they rely entirely on the underlying \ac{KG}, which may contain errors or be incomplete; thus, they cannot be used to assess the accuracy of the KG itself.%

\paragraph{\textbf{(2) External Evidence-Based Fact Checking}}
DeFacto~\cite{GERBER201585} is a supervised learning method that validates \ac{KG} triples using evidence retrieved on the Web. To compute an evidence score, this method integrates trustworthiness metrics with textual evidence. 
\citet{10.1145/3269206.3269308} proposed a fact validation method that uses textual evidence from a static reference corpus as external knowledge. They verbalized triples into natural language, queried a search engine to retrieve similar corpus sentences, and then extracted evidence and features from these sentences to estimate each \ac{KG} triple’s confidence with a trained model. Recently, \citet{boylan2024kgvalidatorframeworkautomaticvalidation} introduced KGValidator, a framework for the automatic evaluation of \ac{KG} completion models using \acp{LLM}. KGValidator assesses predicted triples by leveraging multiple sources of context, including the LLM’s internal knowledge, user-provided textual documents, and web resources. \textcolor{black}{In contrast to this methodological contribution, FactCheck focuses on providing the supporting evaluation infrastructure -- i.e., datasets, metrics, and curated evidence corpora -- needed to systematically assess and compare such validation approaches.}

Aligning with prior work that incorporates external sources for fact verification~\cite{GERBER201585, 10.1145/3269206.3269308, boylan2024kgvalidatorframeworkautomaticvalidation}, \approach allows LLMs to employ external evidence retrieved from Web SERPs. Additionally, \approach offers several LLM-based baselines, enabling a comparative evaluation of LLM with external evidence-driven solutions.
%Compared to KGValidator, which assessed \ac{LLM} validation performance on small subsets of KG completion datasets (150 examples per dataset, 750 in total) due to API constraints, \approach performs a larger and more comprehensive evaluation. 
Moreover, \approach assesses LLM performance across three real-world \ac{KG} datasets (13,530 facts) tailored for the task, supported by 2M+ retrieved documents as external evidence.

\subsection{Benchmarks and \textcolor{black}{Datasets}}
CRAG~\cite{NEURIPS2024_1435d2d0} is a \textbf{benchmark} designed to evaluate the effectiveness of \ac{RAG} systems, with a focus on  factual accuracy. It includes $4,409$ Question-Answer pairs spanning five domains and eight question categories. To simulate realistic usage scenarios, CRAG offers mock APIs for web and \ac{KG} searches. The benchmark specifically targets challenges such as answering less popular or rapidly evolving facts, assessing LLM performance across varying levels of entity popularity and temporal relevance. 
\textcolor{black}{While CRAG and FactCheck both utilize RAG, they address fundamentally different problems with distinct evaluation goals. Indeed, FactCheck evaluates KG fact validation, prioritizing accuracy and consistency. CRAG cannot replace FactCheck because high-performing QA models often fail at the strict, granular logic required to validate isolated KG triples.} Additionally, \approach provides detailed information on computational costs and resource efficiency, both aspects not extensively covered by CRAG. Hence, although related, these benchmarks address different aspects of factual verification.

%Evaluation results show that LLMs achieve 34\% accuracy on CRAG, improving to 44\% with RAG integration, while some commercial systems reach up to 63\%.

%In relation to \approach, CRAG serves as a complementary effort addressing a different verification problem. While CRAG focuses on evaluating RAG techniques for answering natural language questions across various domains, \approach targets the verification of facts stored within \acp{KG}. \approach examines whether the structured triples in \acp{KG} accurately represent real-world knowledge. This distinct focus is crucial as \acp{KG} serve as foundational structured knowledge resources that many systems rely on as authoritative sources. 

Beyond CRAG, there are several pipelines and shared tasks for fact-checking purposes targeting textual claims.
RumourEval~\cite{gorrell-etal-2019-semeval} evaluated classification systems by analyzing social media posts by stance detection and rumor veracity verification, employing a dataset containing data from Twitter and Reddit. CLEF CheckThat!~\cite{alam2025clef2025checkthatlabsubjectivity} offers sentence-level subjectivity detection in news articles.
ClaimBuster~\cite{arslan2020benchmark} introduced an automated end-to-end fact-checking pipeline integrating claim detection, matching, and verification. As said, these benchmarks primarily target unstructured textual claims and cannot be used for KG fact verification.

Few \textbf{datasets} have been proposed for KG verification~\cite{GERBER201585,ojha-talukdar-2017-kgeval,Marchesin_Silvello_Alonso_2024}. A key one is \textit{FactBench}~\cite{GERBER201585}, built from DBpedia~\cite{dbpedia-large-scale} and Freebase~\cite{freebase} \acp{KG} to evaluate validation systems on systematic errors. {Other datasets include \textit{YAGO}~\cite{ojha-talukdar-2017-kgeval} and \textit{DBpedia}~\cite{Marchesin_Silvello_Alonso_2024}, %and \textit{NELL}~\cite{ojha-talukdar-2017-kgeval}, on the other hand, 
which consist of samples drawn from their respective KGs and manually annotated by experts for correctness.} While these datasets have been employed in both manual and automated verification settings, they have seen minimal to no use with LLM-based approaches. Hence, we employ FactBench, YAGO, and DBpedia in \approach, as they capture complementary aspects of fact verification challenges, enabling a multifaceted evaluation of LLM-based strategies. Another related dataset is FactKG~\cite{kim2023factkgfactverificationreasoning}, designed for fact verification over \acp{KG}. 
However, FactKG uses \acp{KG} to verify textual claims, whereas our work takes the opposite direction: using external evidence to help LLMs validate KG facts. %For this reason, we do not include FactKG in \approach.

\begin{figure*}[t!]
    \centering
    \includegraphics[width=\linewidth]{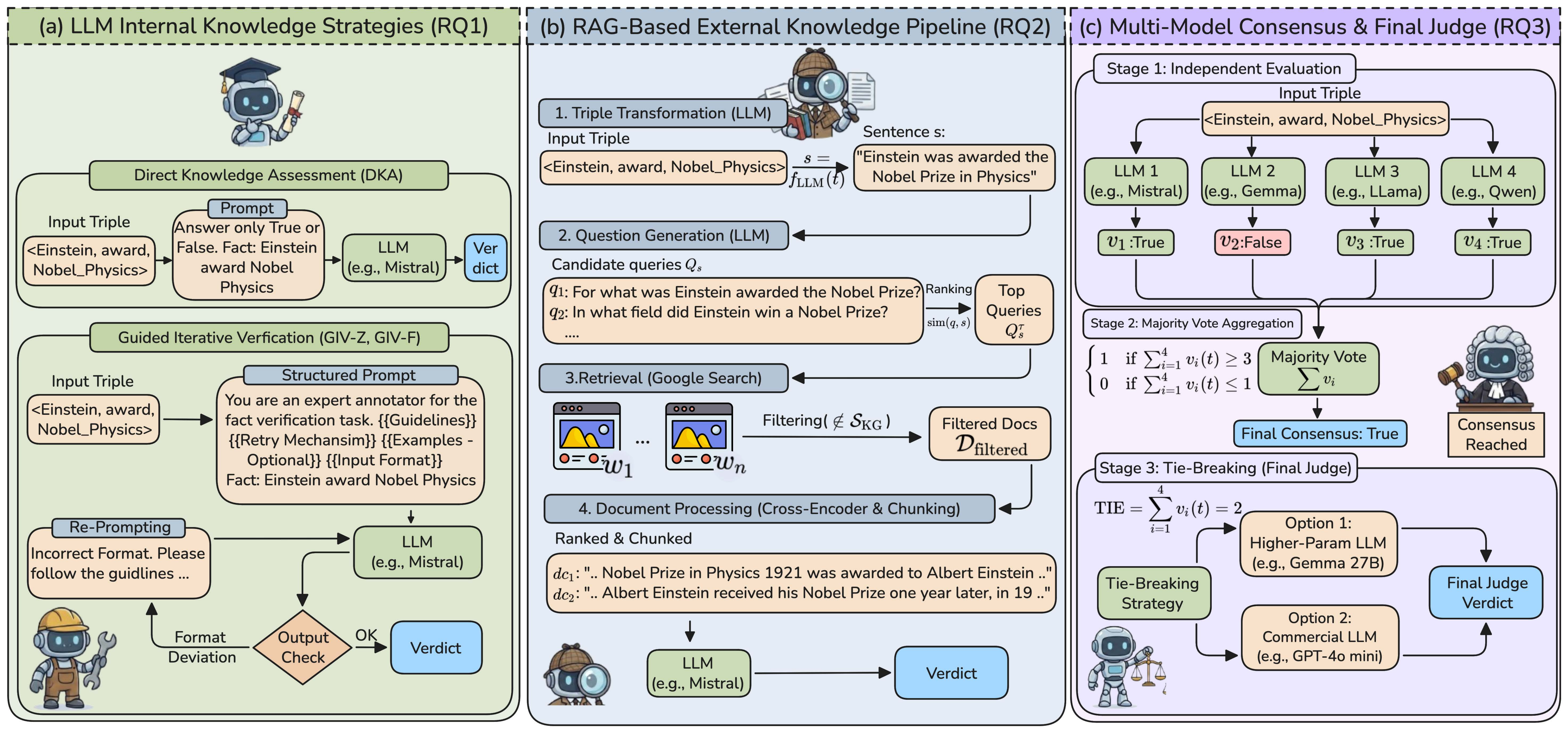}
    \caption{\textcolor{black}{Overall overview of the \approach benchmark.}}
  \label{fig:FactCheckMethods}
\end{figure*}

\section{FactCheck}\label{sec:approach}
%\approach is a benchmark designed to evaluate \acp{LLM} on KG fact-checking tasks. It is built on real-world facts sourced from three \ac{KG} datasets: FactBench, YAGO, and DBpedia. 
This section details the strategies used in \approach to address the study’s \acp{RQ}. The benchmark includes multiple strategies using both open-source and commercial \acp{LLM}. In \S\ref{section:lik}, we present two approaches that rely solely on \acp{LLM}’ internal knowledge to verify \ac{KG} facts (RQ1). In \S\ref{subsec:ek}, we introduce a \ac{RAG} approach that augments \acp{LLM} with external evidence (RQ2). Finally, \S\ref{subsec:mmc} describes a multi-model consensus strategy that aggregates predictions from multiple \acp{LLM} to improve verification accuracy (RQ3). %All prompts and methods are available in the shared GitHub repository.

\subsection{LLM Internal Knowledge}\label{section:lik}
To address \ac{RQ}1, \approach employs two different strategies:

\textbf{\acf{DKA}} is a simple strategy consisting of a basic, direct prompt for the LLM without any further guidance. % or complex instructions.
DKA aims to evaluate the ability of LLMs to verify facts using only internal knowledge. % embedded during their training process.
We use DKA as the baseline for comparing different \acp{LLM} and more advanced strategies. \textcolor{black}{An example is reported in the top left part of Figure~\ref{fig:FactCheckMethods}(a)}.

\textbf{\acf{GIV}} \textcolor{black}{(see the bottom left part of Figure~\ref{fig:FactCheckMethods}(a))} is an iterative prompting approach leveraging a structured prompt template that outlines the expected output format, and, optionally, enforces dataset-specific constraints. If a model's output is non-conformant, the system triggers a re-prompting, explicitly flagging the non-compliance. Responses that repeatedly fail to meet the criteria are marked as invalid. We consider both zero- and few-shot settings. \textcolor{black}{In the few-shot setting, we include a small set of correctly evaluated triples as examples to guide the model's understanding of the task. These examples are shared across datasets and KG-independent at the semantic level, while their encoding is adapted to the target KG to align with predicate and schema conventions.} %In the few-shot setting, we include a small number of correctly evaluated triples as examples to guide the model's understanding of the task. %In the zero-shot setting, no such examples are provided. 

\subsection{External Knowledge}\label{subsec:ek}
To address RQ2, we enhance LLMs with RAG. % by feeding the LLM with external knowledge relevant to the triple under evaluation. 
Given a KG triple $t$, we retrieve a set of documents $\mathcal{D}$ containing potentially supporting or refuting evidence. We implement this through a multistage pipeline comprising four main phases: (1) triple transformation, (2) question generation and ranking, (3) document retrieval and filtering, and (4) document processing and chunking. \textcolor{black}{Figure~\ref{fig:FactCheckMethods}(b) illustrates the core components of the RAG-based verification engine in \approach.}

\textcolor{black}{In the \textbf{Triple Transformation} phase (1), structured \ac{KG} triples are converted into human-readable sentences. This transformation is performed using an \ac{LLM} to address the substantial variability in how different \acp{KG} represent $\langle S, P, O \rangle$ data. \acp{KG} follow heterogeneous conventions for encoding triples, and these source-specific formats often hinder effective information retrieval. Common issues include (1) KG-specific namespaces (e.g., \textit{dbpedia.org/resource/:term:}); (2) special notation such as underscores or camelCase (e.g., \textit{isMarriedTo}, \textit{Alexander\_III\_of\_Russia}); and (3) predicates that lack sufficient grammatical or semantic context. Such representations can restrict search results to the original source pages from which the triples were extracted, thereby introducing retrieval bias and limiting coverage during evaluation. By contrast, natural language reformulations facilitate the discovery of a broader range of relevant web sources. %, which is essential for downstream validation steps. 
We define this process as a transformation function $s = f_{\text{\scriptsize LLM}}(t)$ that maps a triple $t$ to a natural language sentence $s$.}

In the \textbf{Question Generation and Ranking} phase (2), for any given sentence $s$, we prompt an LLM to generate a set of candidate queries $\mathcal{Q}_s = \{q_1, q_2, \ldots, q_{k_q}\}$. The goal of generating multiple questions is to broaden the semantic coverage of the original triple, improving the chances of retrieving relevant evidence -- even when the input is ambiguous, noisy, or underspecified. \textcolor{black}{Generating multiple questions also helps mitigate the paraphrasing bias that the LLM may introduce when turning triples into natural language. By formulating several distinct questions, we broaden the range of possible interpretations of a given triple, thereby weakening the link to any single facet that might otherwise be imposed by one particular LLM-generated paraphrase.}
To identify the most informative queries, we apply a cross-encoder model \textcolor{black}{(\textit{jina-reranker-v1-turbo-en}), which corresponds to the normalized dot product between the cross-encoder's final representation and a learned relevance vector (i.e., a sigmoid-scaled dot-product score). This score reflects the semantic proximity between a candidate query $q \in \mathcal{Q}_s$ and the original sentence $s$.} The resulting set is $\mathcal{Q}_s^{\text{ranked}} = \{q_{(1)}, q_{(2)}, \ldots, q_{(k_q)}\}$, where $\mathrm{sim}(q_{(i)}, s) \geq \mathrm{sim}(q_{(i+1)}, s)$ for all $i \in \{1, 2, \ldots, k_q - 1\}$. We retain the top-$\tau$ queries, denoted as $\mathcal{Q}_s^{\tau}$, using a predefined threshold $\tau \in [0,1]$ to ensure only the most relevant queries are used.

In the \textbf{Document Retrieval and Filtering} phase (3), we issue each query in $\mathcal{Q}_s^{\tau}$ to Google Search using \textcolor{black}{specific parameters to ensure consistency. We set lr = ``lang\_en'' and hl = ``en'' to enforce English content and interface settings, and gl = ``us'' to standardize the geolocation to the United States, thereby mitigating local personalization bias. Using num = ``100'',} we collect the top $n_{\max} = 100$ retrieved webpages, denoted as $\mathcal{R}(q) = \{w_1, w_2, \ldots, w_{n_{\max}}\}$.
For each webpage $w_i \in \mathcal{R}(q)$, we extract its textual content, denoted as $\text{text}(w_i)$. The set of documents retrieved for a given query $q$ is then defined as $\mathcal{D}(q) = \{d_i = \text{text}(w_i) \mid w_i \in \mathcal{R}(q)\}$. To obtain the full document pool associated with the original triple $t$, we take the union over all queries in $\mathcal{Q}_s^{\tau}$: $\mathcal{D} = \bigcup_{q \in \mathcal{Q}_s^{\tau}} \mathcal{D}(q)$. To ensure evidence independence and avoid circular verification, we define $\mathcal{S}_{\text{KG}}$ as the set of original \ac{KG} sources -- for instance, Wikipedia entries when verifying facts from DBpedia and FactBench datasets. We use this set to filter out any retrieved documents that \textcolor{black}{directly} originate from these sources. The resulting filtered document set is defined as $\mathcal{D}_{\text{filtered}} = \{d \in \mathcal{D} \mid \text{source}(d) \notin \mathcal{S}_{\text{KG}}\}$. %This step ensures that the evidence used for verification does not inadvertently rely on the same KG data being evaluated. 

Finally, in the \textbf{Document Processing and Chunking} phase (4), we use  a cross-encoder to identify the $k_d$ most relevant documents with respect to the sentence $s$. % generated from the triple $t$. 
For each document $d \in \mathcal{D}_{\text{filtered}}$, a similarity score $\text{sim}_d(d, s)$ is computed using the same approach as above. The top $k_d$ documents, ranked by similarity, form the final set $\mathcal{D}_{\text{final}} = \{d_{1}, d_{2}, \ldots, d_{k_d}\}$. Each document in $\mathcal{D}_{\text{final}}$ is segmented into smaller, overlapping passages using a sliding window chunking strategy. These chunks are subsequently used as contextual input in the LLM prompt during the fact validation stage. %All the relevant parameters are reported in Table \ref{tab:system-configurations}.

%embedded using an embedding model to generate semantic representations, which are 

\subsection{Multi-Model Consensus}\label{subsec:mmc}

% \begin{figure}[t!]
%     \centering
%     \includegraphics[width=\linewidth]{{figures/Consensus}}
%   \caption{Multi-model consensus methodology for \ac{KG} fact validation. In Stage 1, multiple open-source LLMs independently evaluate the same fact. Stage 2 implements a Majority Vote mechanism that aggregates these individual model decisions to form a consensus and final verification. If the vote results in a tie, a final-judge component-typically a higher-parameter model or commercial LLM-(Stage 3) resolves the ambiguity and determines the ultimate verification outcome.}
%   \Description{Overview of the FactCheck System.}
%   \label{fig:Consensus}
% \end{figure}

Since \acp{LLM} can output different answers for the same fact-checking task, we also explore a model consensus strategy \textcolor{black}{(Figure \ref{fig:FactCheckMethods}(c))}. Building on \S\ref{section:lik} and \S\ref{subsec:ek}, let $\mathcal{M} = \{M_1, M_2, M_3, M_4\}$ be the set of \acp{LLM}. For each triple $t$, each model $M_i \in \mathcal{M}$ produces a binary verdict $v_i(t) \in \{0, 1\}$, where 0 means ``false'' and 1 means ``true''.

We employ a simple majority vote strategy to determine the final verdict. The consensus decision $V_{\text{final}}(t)$ for a given triple $t$ is:
$$\small V_{\text{final}}(t) = \begin{cases}
1 & \text{if } \sum_{i=1}^{4} v_i(t) \geq 3 \\
tie & \text{if } \sum_{i=1}^{4} v_i(t) = 2  \\ 
0 & \text{otherwise}
\end{cases}$$
The strategy aims to mitigate errors by reducing the impact of outlier predictions. In the event of a tie, we apply a conflict resolution strategy. Let $M_{\text{judge}}$ be the \textit{final judge} module responsible for breaking ties. We explore two approaches for defining $M_{\text{judge}}$:

\begin{itemize}
[noitemsep,topsep=0pt,parsep=0pt,partopsep=0pt,leftmargin=18pt]
\item[(1)] A higher-parameter variant of one of the models in our set $\mathcal{M}$, selected based on its consistency score \textcolor{black}{$\text{CA}_M$}. This score represents the proportion of instances where the model's output agrees with the majority prediction across datasets -- serving as a proxy for its alignment with correct outcomes. We test both the most consistent (highest \textcolor{black}{$\text{CA}_M$}) and least consistent (lowest \textcolor{black}{$\text{CA}_M$)} models, upgrading them to higher-parameter versions (e.g., Gemma2:9B $\rightarrow$ 27B).

\item[(2)] A commercial model with a different architecture and training pipeline -- such as \textit{GPT-4o mini} -- to offer an independent perspective in resolving ambiguous cases. 
\end{itemize}

\section{Benchmark Construction}\label{sec:construction}
In this section, we present the entire pipeline for constructing the \approach benchmark. First, in \S\ref{subsec:datasets}, we detail the process of collecting triples from existing KG datasets, along with the creation of a new dataset specifically tailored for the \ac{RAG} methodology. 
Next, in \S\ref{subsec:models} and \S\ref{subsec:performance_metrics}, we describe the \acp{LLM}, the evaluation metrics, and the automated assessment procedures used in \approach.

\subsection{Datasets}\label{subsec:datasets}
The \approach dataset consists of two main components: (i) triples derived from three real-world \acp{KG}, and (ii) content retrieved from Google \acp{SERP}. This section describes each of these components and introduces the mock API, which mimics a realistic scenario and provides standardized access to the dataset for reproducible experimentation.

\paragraph{\textbf{KG Datasets.}}
We include triples from three real-world and widely used KG datasets -- FactBench, YAGO, and DBpedia. \textcolor{black}{Note that we employ these datasets with a snapshot-based semantics: a triple is deemed true if it is supported by the underlying KG snapshot used to build it, and false otherwise.} Table~\ref{tab:freq} summarizes the key statistics for each of these datasets.
\begin{table}[ht]
\small
    \caption{Summary of FactBench, YAGO, and DBpedia datasets.}
  \label{tab:freq}
  \begin{tabular}{lccc}
    \toprule
    & \textbf{FactBench} & \textbf{YAGO} & \textbf{DBpedia}\\
    \midrule
        Num. of Facts & 2,800 & 1,386 & 9,344 \\
        Num. of Predicates & 10 & 16 & 1,092 \\
        Avg. Facts per Entity & 2.42 & 1.69 & 3.18 \\
        Gold Accuracy ($\mu$) & 0.54 & 0.99 & 0.85 \\
  \bottomrule
\end{tabular}
\end{table}

\textbf{FactBench} is a multilingual benchmark developed by \citet{GERBER201585} to evaluate fact validation algorithms. It includes ten relation types and supports English, German, and French. In \approach, we focus exclusively on the English subset. Positive (correct) facts are sourced from DBpedia and Freebase, while negative (incorrect) facts are generated systematically by altering the correct ones -- ensuring adherence to domain and range constraints. We use a configuration with a \textcolor{black}{proportion of positive facts} of $\mu = 0.54$, achieved by mixing correct facts with incorrect ones generated through various negative sampling strategies~\cite{marchesin_silvello-2025}.

\textbf{YAGO} is an evaluation dataset sampled from the YAGO KG, originally introduced by \citet{ojha-talukdar-2017-kgeval} and widely adopted for KG accuracy estimation~\cite{gao_etal-2019,marchesin_silvello-2024,marchesin_silvello-2025}. It comprises $1,386$ facts spanning 16 distinct predicates, with an average of $1.69$ facts per entity. All facts are annotated by crowdworkers, resulting in a gold standard accuracy of $\mu = 0.99$. This high accuracy presents a unique challenge for fact-checking, as LLMs may be biased toward classifying all facts as correct, thereby inflating performance metrics. 

\textbf{DBpedia} is an evaluation dataset sampled from the DBpedia KG, originally introduced by \citet{Marchesin_Silvello_Alonso_2024}. It was constructed using a combination of sampling and active learning techniques, with both expert and layman annotators involved to ensure high annotation quality. The triples were acquired from the 2015-10 English version of DBpedia, with subject entities required to be part of triples that include \texttt{rdfs:label} and \texttt{rdfs:comment} predicates. To focus exclusively on factual assertions, T-Box triples -- those representing ontological entities and schema-level relationships -- were excluded, retaining only A-Box assertions, which represent concrete factual claims. Each triple was annotated by at least three annotators, resulting in a dataset of $9,934$ triples with a gold standard accuracy of $\mu = 0.85$, covering $1,092$ distinct predicates.

\paragraph{\textbf{RAG Dataset.}}
We constructed a \ac{RAG} dataset comprising questions derived from \ac{KG} facts and corresponding search results. This dataset was created as support to effectively evaluate \ac{LLM} performance in fact validation tasks involving external knowledge. The dataset consists of two main components: the generated questions and their associated search results obtained from Google \acp{SERP}.

For \textbf{Questions}, we used an \ac{LLM} to generate $k_q=10$ distinct questions for each transformed triple $s$, aiming to explore different facets of the underlying fact. For dataset construction, we included all questions that were successfully extracted from the model's output. Each question is published along with its corresponding similarity score, computed with respect to the transformed triple. \approach comprises a total of $Q = 130,820$ questions generated for $13,530$ facts. Each fact is associated with a variable number of questions ($q_t$) ranging from $\min(q_t) = 2$ to $\max(q_t) = 10$, with a mean of $\mu_{q_t} = 9.67$ and a median of $\tilde{q}_t = 10.00$. %, indicating that the vast majority of facts have the maximum number of associated questions.

Each question is assigned a similarity score $\delta \in [0,1]$ that quantifies its semantic closeness to the transformed triple. Across all questions, the similarity scores exhibit a mean of $\mu_\delta = 0.63$ and a median of $\tilde{\delta} = 0.66$. The standard deviation is $\sigma_\delta = 0.25$, indicating moderate variability. The first quartile is $Q_1 = 0.44$ and the third is $Q_3 = 0.84$, resulting in an \ac{IQR} of $\text{IQR} = Q_3 - Q_1 = 0.40$, which confirms substantial variation in similarity scores across the dataset.

To further analyze this distribution, we categorize the questions into three similarity tiers: high similarity ($\delta \geq 0.70$), constituting $45\%$ of the dataset; medium similarity ($0.40 \leq \delta < 0.70$), accounting for $34\%$; and low similarity ($\delta < 0.40$), making up the remaining $21\%$. This distribution shows that $79\%$ of the dataset consists of questions with at least moderate similarity to the transformed triple ($\delta \geq 0.40$), and nearly half show high similarity. This range of similarity levels covers both semantically close and more loosely related interpretations of each fact.

Regarding \textbf{Google Search Results}, for each fact, we submitted the transformed original triple along with the top three generated questions -- ranked by their similarity scores -- to Google Search. After parsing the HTML responses, we retrieved each URL using the \textit{GRequests} Python library. The content of the resulting webpages was extracted using the \textit{newspaper4k} \footnote{\url{https://newspaper4k.readthedocs.io}} Python package. 

The corpus consists of $D = 2{,}090{,}305$ documents across $13{,}530$ triples. Each triple $t$ is linked to $d_t$ documents, with $\min(d_t) = 0$, $\max(d_t) = 337$, mean $\mu_{d_t} = 154.51$, and median $\tilde{d}_t = 160$. The slightly higher median indicates a mild negative skew, with most triples having document counts around or just above the mean.

We define $\mathcal{E}_{\text{text}} \subset D$ as the subset of documents with empty text content. This subset contains $|\mathcal{E}_{\text{text}}| = 263,515$ documents, representing the $13\%$ of the entire collection. Consequently, the text coverage rate -- i.e., the proportion of documents presenting text content -- is $1 - |\mathcal{E}_{\text{text}}|/|D| = 0.87$ ($87\%$). This high coverage rate supports the reliability of the constructed document collection. 

\begin{table}[t!]
    \small
    \centering
    \caption{Summary of average time and token usage for each step in the RAG dataset generation pipeline.}
        \begin{tabular}{lcc}
            \toprule
            \textbf{Task} & \textbf{Avg. Time} & \textbf{Avg. tokens} \\
            \midrule
            % Human understandable text & 1.3164 sec & 343.16 \\
            Question Generation & 9.60 sec & 672.58 \\
            \hline
            % Sorting the questions by similarity~\ref{sss:relevance} & 0.013 sec & --\\
            Get documents (Google pages) & 3.60 sec & -- \\
            Fetch documents for each triple & 350 sec & -- \\
            \bottomrule
        \end{tabular}
    \label{tab:pipeline-performance-report}
\end{table}

In Table \ref{tab:pipeline-performance-report}, we report the time consumption and token expenditure incurred during the generation of the \ac{RAG} dataset.
Overall, question generation requires an average of 9.60 seconds per fact, whereas the complete Google results retrieval process takes approximately $364.4$ seconds. %$\theta_{\text{tot}} = \theta_{\text{base}} + \theta_{\text{retrieval}} = (3.60 \times 4)+ 350 = 364.4$ seconds. The factor of 4 reflects the number of queries submitted to Google Search per fact: the transformed original triple and the three top-ranked generated questions.

To ensure fairness and reproducibility in evaluation, we generated all questions and collected the corresponding Google \ac{SERP} results in advance. This provides a consistent evidence base for \acp{LLM}, avoiding discrepancies caused by changes in live search outputs. The complete dataset is publicly available on our HuggingFace project page and accessible via the mock API.\footnote{\url{https://huggingface.co/datasets/FactCheck-AI/FactCheck}}

\paragraph{\textbf{Mock API}}
In \approach, we integrate a web search-like API for content retrieval to simulate realistic scenarios for \ac{RAG}. This API facilitates reproducible benchmarking by offering standardized access to pre-collected search data, thereby removing temporal variability in search results.

For each fact in the considered datasets, we issued queries using both the transformed triple and the top three generated questions. We stored the first 100 results for each query from Google \ac{SERP}, and subsequently retrieved and preserved the actual content of each linked webpage. As previously discussed, we filtered out sources directly related to the original fact to avoid circular verification.

We implemented standardized endpoints that emulate conventional web search APIs while returning consistent results from our dataset. Through this mock API, researchers can perform identical retrieval operations across multiple experimental runs, ensuring fair comparisons between different \ac{LLM} configurations, prompting strategies, and verification approaches. The mock API can be accessed at \url{https://factcheck-api.dei.unipd.it/}. Full documentation is available on GitHub.\footnote{\url{https://github.com/FactCheck-AI/FactCheck-MockAPI}}

\subsection{Models}\label{subsec:models}
We integrate four open-source LLMs in the 7-9B parameter range as the backbone of our KG fact validation pipeline: Gemma2, Qwen2.5, Mistral, and Llama3.1. \textcolor{black}{We prioritize open-source models for several reasons. First, they can be deployed in diverse environments, including settings with strict data privacy requirements or limited API access, as they can be hosted locally without relying on external services. Second, they offer greater tunability, allowing fine-tuning on domain-specific data or adaptation to specialized fact validation tasks. Third, they are significantly more cost-effective for large-scale applications, avoiding per-token API costs that can become prohibitive when processing extensive \acp{KG}. %Furthermore, the selected models demonstrate strong linguistic capabilities, reasoning skills, and compatibility with \ac{RAG}, especially in low-resource environments. 
To provide a performance reference and assess the gap between open-source and commercial solutions, we also include GPT-4o mini, a commercial model from OpenAI.}

% PREVIOUS VERSION
% These models were selected for their strong linguistic capabilities, reasoning skills, and compatibility with \ac{RAG}. Furthermore, adopting smaller-scale LLMs enhances benchmark usability, especially in low-resource environments.

\textbf{Gemma2:9B}, developed by Google, is an open-source 9B parameter model optimized for efficiency~\cite{gemmateam2024gemma2improvingopen}, excelling in natural language understanding and generation. %, supporting an 8k token context window.

\textbf{Qwen2.5:7B}, from Alibaba Cloud, is an open-source 7B parameter model notable for improved instruction-following, reasoning, and structured data handling~\cite{qwen2.5,qwen2}.

\textbf{LLaMA3.1:8B}, by Meta, is an open-source 8B parameter model that features an extensive 128k token context window and enhanced multilingual support, making it suitable for long-context and diverse language tasks~\cite{dubey2024llama3herdmodels}. 

\textbf{Mistral:7B}, developed by Mistral AI, is a 7B parameter model known for its performance and compactness balance, demonstrated across various benchmarks~\cite{jiang2023mistral7b}.

\textbf{GPT-4o mini}, developed by OpenAI as a smaller variant of GPT-4o, offers strong reasoning capabilities with reduced latency and cost~\cite{openai2024gpt4omini}, serving as a commercial baseline for advanced knowledge retrieval and fact verification.

\subsection{Performance Metrics and Evaluation}\label{subsec:performance_metrics}
To assess the effectiveness of the considered fact validation strategies, we focus on two key measures: \textcolor{black}{Class-wise F1 Score and Consensus Alignment.} These measures are chosen to account for class imbalance, capture per-class performance, and evaluate agreement for multi-model consensus approaches. We also evaluate efficiency by computing the average response time required by each considered strategy to provide a verification response.

% \textcolor{red}{\textbf{Balanced Accuracy} (BAcc) is defined as the average of sensitivity and specificity, offering a robust evaluation in imbalanced classification settings. It is computed as:
% \begin{equation*}
%     \text{BAcc} = \frac{1}{2}\left(\frac{\text{TP}}{\text{TP}+\text{FN}} + \frac{\text{TN}}{\text{TN}+\text{FP}}\right)
% \end{equation*}
% where TP, FN, TN, and FP denote true positives, false negatives, true negatives, and false positives, respectively.}

% \textbf{Macro-averaged F1 Score} ($F_{1\text{macro}}$) captures the unweighted mean of the F1 score computed independently for each class. This ensures that performance on minority classes is treated equally to that on majority classes. It is defined as:
% \begin{equation*}
%     \text{F1}_{\text{macro}} = \frac{1}{N} \sum_{i=1}^{N} \left( \frac{2 \cdot \text{Precision}_i \cdot \text{Recall}_i}{\text{Precision}_i + \text{Recall}_i} \right)
% \end{equation*}
% where $N$ is the number of classes, and $\text{Precision}_i$ and $\text{Recall}_i$ are the precision and recall for class $i$.

\textcolor{black}{\textbf{Class-wise F1 Scores} ($F1(c)$) are calculated independently for ``True'' ($T$) and ``False'' ($F$) labels to assess performance on each single category, rather than aggregating them. This granular view highlights potential disparities in model performance between the two classes. The $F1$ score for a given class $c \in \{T, F\}$ is defined as:
$$F1(c) = \frac{2 \cdot \text{Precision}(c) \cdot \text{Recall}(c)}{\text{Precision}(c) + \text{Recall}(c)},$$ 
where $\text{Precision}(c)$
and $\text{Recall}(c)$ denote the precision and recall calculated specifically for class $c$.}

% \textcolor{red}{\textbf{Consistency} ($\text{Cons}_M$) measures the agreement between a given model’s predictions and the majority vote across all evaluated facts. Specifically, for a model $M$, it is defined as:
% \begin{equation*}
%     \text{Cons}_M = \frac{|\{ t \in G \mid \operatorname{response}(M, t) = \operatorname{majorityVote}(t) \}|}{|G|}
% \end{equation*}
% where $G$ denotes the set of \ac{KG} dataset triples, $\operatorname{response}(M, t)$ the prediction of model $M$ for triple $t$, and $\operatorname{majorityVote}(t)$ the \textit{consensus} label assigned by the majority of models. A consistency of $1.0$ indicates perfect alignment with the majority, whereas a value of $0.0$ indicates complete disagreement. For instance, a score of $0.7$ implies that the model agrees with the majority in $70\%$ of the cases.}

\textcolor{black}{\textbf{Consensus Alignment} ($\text{CA}_M$) quantifies the agreement between a given model’s predictions and the majority vote across all evaluated facts. Specifically, for a model $M$, it is defined as:
\begin{equation*}
    \text{CA}_M = \frac{1}{|G|} \sum_{t \in G} \mathbb{I}(\operatorname{response}(M, t) = \operatorname{majorityVote}(t))
\end{equation*}
where $\mathbb{I}(\cdot)$ denotes the indicator function, which evaluates to $1$ if the condition is met and $0$ otherwise. Here, $\operatorname{response}(M, t)$ represents the prediction of model $M$ for triple $t$, and $\operatorname{majorityVote}(t)$ is the label assigned by the majority of models in the ensemble.
The $\text{CA}_M$ score ranges from $0$ to $1$.
High $\text{CA}_M$ identifies the ``Most Representative'' model serving as the best single proxy for the group's consensus, and low $\text{CA}_M$ identifies the ``Outlier'' model. This indicates a model that systematically deviates from the majority opinion.
}

To evaluate \textbf{efficiency}, we measure the fact average response time in seconds, denoted as $\bar{\theta}$. To ensure a robust assessment that is not distorted by extreme values, we apply an outlier removal process based on the \ac{IQR} method. Given a model-dataset pair, let $\Theta = \{ \theta_1, \theta_2, \ldots, \theta_n \}$ be the set of model's response times over the $n$ dataset facts. We start by computing the first $Q_1 = P_{25}(\Theta)$ and third $Q_3 = P_{75}(\Theta)$ quartiles, and then derive $\text{IQR} = Q_3 - Q_1$. Finally, we define the lower and upper bounds for acceptable values as $L_{\text{lower}} = Q_1 - 1.5 \times \text{IQR}$ and $L_{\text{upper}} = Q_3 + 1.5 \times \text{IQR}$. We exclude all response times outside these bounds, resulting in the filtered set $\Theta' = \{\theta \in \Theta \mid L_{\text{lower}} \leq \theta \leq L_{\text{upper}}\}$. The average response time per fact is then the mean response time over the filtered set, computed as: $\bar{\theta} = \frac{1}{|\Theta'|}\sum_{\theta \in \Theta'} \theta$.

%For the Efficiency Measure ($T_f$), we calculate the average response time by removing outliers to ensure robust performance assessment. For each dataset and model combination, given a set of duration measurements $T = \{t_1, t_2, \ldots, t_n\}$, we apply the \ac{IQR} method. We compute $Q_1 = P_{25}(T)$ (first quartile) and $Q_3 = P_{75}(T)$ (third quartile), then calculate $IQR = Q_3 - Q_1$. We define the lower and upper bounds for the detection of outliers as $L_{lower} = Q_1 - 1.5 \times IQR$ and $L_{upper} = Q_3 + 1.5 \times IQR$. The filtered set of measurements excluding outliers is then $T' = \{t \in T \mid L_{lower} \leq t \leq L_{upper}\}$. The refined average response time $\overline{T}_f$ for each fact $f$ is calculated as $\overline{T}_f = \frac{1}{|T'|}\sum_{t \in T'} t$.

\section{Experimental Setup}\label{sec:setup}
This section details the technical specifications, computational infrastructure, and methodological framework used to implement \approach. We describe the hardware environments, model configurations, and procedural protocols. % that ensure consistent evaluation across all experiments and datasets.

%We outline the technical specifications and operational frameworks to implement \approach. We describe the computational infrastructure, model parameterization, and procedural protocols established for consistent evaluation across different methodologies and datasets.

%The experimental framework of \approach requires a precise specification of hardware resources, software components, and parameter settings to ensure reproducibility. 
%In \S\ref{pargraph:hardware_software}, we describe the technical infrastructure (including both hardware and software) used to perform the benchmark. Then, in Paragraphs \ref{pargraph:RAG} and \ref{pargraph:multi}, we report all the different configuration parameters were systematically or manually selected to perform empirical performance evaluations. These configurations are selected with particular attention to resource allocation optimization for \ac{LLM} inference tasks, particularly within \ac{RAG} method.

To retrieve Google \ac{SERP} results, we employed a Unix-based server equipped with 2 CPU cores and 4 GB of RAM. For triple transformation and question generation, we used a MacBook Pro powered by an Apple M2 Max chip with 32 GB of RAM. All other experiments involving \acp{LLM}, including prompting and evaluation, were conducted on a Mac Studio (Model: Mac14,14) equipped with an Apple M2 Ultra chip featuring 24 cores (16 performance and 8 efficiency cores) and 192 GB of unified memory.

%To fetch Google \ac{SERP}, we used unix-based server with 2 cores and 4GB of RAM. We use a Macbook Pro with an M2 max chip and 32GB of RAM for triple transformation and question generation. All other tests to prompt \acp{LLM} are performed in a Mac Studio (Model: Mac14,14) equipped with an Apple M2 Ultra chip featuring 24 cores (16 performance, 8 efficiency) and 192 GB of memory.

Open-source \acp{LLM} were executed locally using \textit{Ollama},\footnote{\url{https://ollama.com/}} an open-source framework that streamlines the deployment and usage of \acp{LLM} on local machines. For monitoring model behavior, including token usage and inference time, we integrated OpenTelemetry via tooling from the OpenLIT project.\footnote{\url{https://openlit.io/}} This setup provides robust monitoring for \acp{LLM}, vector databases, and GPUs usage.

%For running open-source \acp{LLM}, we use \textit{Ollama}\footnote{\url{https://ollama.com/}}. Ollama is an open source project that simplifies the process of setting up, running, and using \acp{LLM} locally on your machine. To monitor tokens and time \acp{LLM}, we use OpenTelemetry observability of the OpenLIT~\footnote{\url{https://openlit.io/}} team. This tool offers monitoring ability for \acp{LLM}, vector databases, and GPUs.

% \subsubsection{\textbf{Prompting Technique}}\label{pargraph:Prompting}
% \textcolor{red}{variants for prompt technique, the number of iterations and so on, Andrea's paper \ldots}

\begin{table}[t!]
    \small
    \caption{Configuration parameters used in the RAG pipeline.}
    \begin{tabular}{ll}
    \toprule
        \textbf{RAG Component} & \textbf{Parameter}  \\
        \midrule
        Human Understandable Text & Gemma2:9b  \\
        \midrule
        Question Generation & Gemma2:9b  \\
        Question Relevance & Jina-reranker-v1-turbo-en  \\
        Relevance Threshold & 0.5 \\
        Selected Questions & 3  \\
        \midrule
        Selected Documents ($k_d$) & 10 \\
        \midrule
        Document Selection & ms-marco-MiniLM-L-6-v2  \\
        Embedding Model & bge-small-en-v1.5  \\
        Chunking Strategy & Sliding Window (size = 3)\\
        % Similarity Cut-off & Simple  \\
        % Cut-off Threshold & 0.3 \\
        % Top\_k & 6  \\
        \bottomrule
    \end{tabular}
    \label{tab:system-configurations}
\end{table}

Configuration parameters for the \ac{RAG} pipeline are reported in Table~\ref{tab:system-configurations}. These settings were determined through a series of experiments comparing alternative configurations. The results of these ablation studies are available in the GitHub repository.\footnote{\url{https://github.com/FactCheck-AI/FactCheck/blob/main/extra-experiments/ablation_study_results/README.md}}

For multi-model consensus, we have two distinct experimental scenarios: one using higher-parameter open-source models, and the other using a commercial LLM, as described in \S\ref{subsec:mmc}. In the open-source scenario, after computing model consistency across datasets, we selected the models with the highest and lowest consistency scores. We then replaced the base versions with their larger counterparts: LLaMA3.1 (8B $\rightarrow$ 70B), Gemma2 (9B $\rightarrow$ 27B), Qwen2.5 (7B $\rightarrow$ 14B), and Mistral (7B $\rightarrow$ \texttt{nemo:12B}). In the commercial baseline scenario, we used OpenAI GPT-4o mini, providing a strong reference point for comparison with open-source alternatives.

\begin{table*}[t!]
    \small
    \caption{\textcolor{black}{Performance evaluation of fact verification systems. The assessment covers various methodologies (DKA, GIV-Z, GIV-F, RAG). In each column, the best-performing method is highlighted in bold, and the second-best method is underlined.}}
        \resizebox{0.8\linewidth}{!}{\begin{tabular}{lc | cc | cc | cc | cc || cc}
            \toprule
            \multirow{2}{*}{Dataset} & 
            \multirow{2}{*}{Method} & 
            \multicolumn{2}{c|}{Gemma2} & 
            \multicolumn{2}{c|}{Qwen2.5} &
            \multicolumn{2}{c|}{Llama3.1} & 
            \multicolumn{2}{c||}{Mistral} & 
            \multicolumn{2}{c}{GPT-4o mini} \\
            \cmidrule(lr){3-4} 
            \cmidrule(lr){5-6} 
            \cmidrule(lr){7-8} 
            \cmidrule(lr){9-10}
            \cmidrule(lr){11-12} 
            & & 
            $F1(T)$ & $F1(F)$  &
            $F1(T)$ & $F1(F)$  &
            $F1(T)$ & $F1(F)$  &
            $F1(T)$ & $F1(F)$  &
            $F1(T)$ & $F1(F)$  \\
            \midrule
            % \rowcolor{gry}
            \multirow{4}{*}{FactBench} & DKA & 
             0.75  & 0.74 & 
             0.55  & 0.71 &
             0.73  & \underline{0.74} & 
             0.68  & \underline{0.73} &
             \underline{0.52}  & \underline{0.72}   \\
             & GIV-Z & 
             0.73 & 0.73 & 
             0.51 & 0.70  &
             0.52 & 0.70 & 
             0.77 & 0.72  &
             0.48 & 0.71  \\
             & GIV-F & 
            \underline{0.79} & \underline{0.76} & 
            \underline{0.74}  & \underline{0.73} &
            \underline{0.75}  & 0.72 & 
            \underline{0.81}  & \underline{0.73}  &
            0.49  & 0.71   \\
            & RAG & 
            \textbf{0.91} & \textbf{0.89}  & 
            \textbf{0.89} & \textbf{0.85}  &
            \textbf{0.83} & \textbf{0.80}  & 
            \textbf{0.87} & \textbf{0.82}  &
            \textbf{0.91} & \textbf{0.90}  \\ \hline
            \rowcolor{gry}
            \multicolumn{2}{c|}{Mean} & 0.80 & 0.78 & 0.67 & 0.75 & 0.71 & 0.74 & 0.78 & 0.75 & 0.60 & 0.76 \\\hline
            \multirow{4}{*}{YAGO} & DKA & 
            0.82  & 0.02  & 
            0.42  & 0.02  &
            0.71  & 0.02  & 
            0.59  & 0.01  &
            0.48  & 0.02      \\
            & GIV-Z & 
            \underline{0.88} & \textbf{0.03} & 
            0.53 & 0.02 & 
            0.52  & 0.02  & 
            0.75 & \textbf{0.02}  & 
            0.51 & 0.02   \\
            & GIV-F & 
            \textbf{0.92} & 0.02 & 
            \underline{0.72} & \textbf{0.03}   & 
            \underline{0.83}  & 0.02  & 
            \underline{0.90}  & 0.01 & 
            \underline{0.53}  & 0.02   \\
            & RAG & 
            \textbf{0.92} & \textbf{0.03}  & 
            \textbf{0.92} & \textbf{0.03}  &
            \textbf{0.91} & 0.02  & 
            \textbf{0.96} & \textbf{0.02}  &
            \textbf{0.89} & 0.02    \\ \hline
            \rowcolor{gry}
            \multicolumn{2}{c|}{Mean} & 0.89 & 0.03 & 0.65 & 0.03 & 0.74 & 0.02 & 0.80 & 0.02 & 0.60 & 0.02 \\\hline
            \multirow{4}{*}{DBpedia} & DKA & 
            \textbf{0.85}  & 0.36  & 
            0.63  & 0.33    &
            \textbf{0.81}  & 0.29   & 
            0.79  & \underline{0.34}  &
            \underline{0.56}  & 0.31    \\
            & GIV-Z & 
            0.81  & \underline{0.37}  & 
            0.63  & 0.33 &
            0.53  & 0.31   & 
            \underline{0.87}  & 0.23  &
            0.48  & \underline{0.31}    \\
            & GIV-F & 
            \textbf{0.85}  & 0.35  & 
            \underline{0.78}  & \underline{0.36} &
            0.69  & \underline{0.32}  & 
            \textbf{0.89}  & 0.20  &
            0.36  & 0.30   \\
            & RAG & 
            0.79 & \textbf{0.38} & 
            \textbf{0.82} & \textbf{0.39} & 
            \underline{0.74} & \textbf{0.33} & 
            0.82 & \textbf{0.38} & 
            \textbf{0.75} & \textbf{0.37}  \\ \hline
            \rowcolor{gry}
            \multicolumn{2}{c|}{Mean} & 0.83 & 0.37 & 0.72 & 0.35 & 0.69 & 0.31 & 0.84 & 0.29 & 0.54 & 0.32 \\
            
            \bottomrule
        \end{tabular}}
    \label{tab:evaluation_results-partition}
\end{table*}

\section{Experimental Analysis}\label{sec:performance_report}
In this section, we present a comprehensive evaluation of \ac{LLM} performance on the \approach benchmark, evaluating their proficiency in \ac{KG} fact validation. Tables \ref{tab:evaluation_results-partition} and \ref{tab:evaluation_results-partition-2} report the $F1$ scores \textcolor{black}{for true and false labels separately} for each model on the FactBench, YAGO, and DBpedia datasets. This analysis is organized around the three key research questions introduced earlier.

\paragraph{\textbf{RQ1.}}%\label{subsec:rq1}
Table~\ref{tab:evaluation_results-partition} provides an overview of the evaluation results concerning the internal knowledge capabilities of \acp{LLM}. The analysis employs three verification paradigms: Direct Knowledge Assessment (DKA), as well as Guided Iterative Verification in both zero-shot (GIV-Z) and few-shot (GIV-F) contexts.

We observe a sensible performance variability across models and datasets. In the FactBench dataset, Gemma2 achieves the \textcolor{black}{robust capabilities across both classes, reaching $0.79$ for $F1(T)$ and $0.76$ for $F1(F)$ in the GIV-F setting.} In contrast, GPT-4o mini shows \textcolor{black}{a distinct performance asymmetry. While its detection of incorrect facts is comparable to other models $F1(F) \approx 0.71$, its ability to verify true facts is consistently lower $F1(T) \approx [0.48, 0.52]$.} This finding challenges the prevailing view that commercial or larger models outperform smaller or open-source counterparts.

% Among the datasets, FactBench appears to be the most favorable for internal knowledge evaluation, \textcolor{black}{as most models maintain a reasonable balance between $F1(T)$ and $F1(F)$.} On the other hand, YAGO proves to be the most challenging due to its very large number of correct facts ($\mu = 0.99$). While achieving relatively low balanced accuracy scores ($0.44$ to $0.64$), the corresponding F1 scores are even lower ($0.22$ to $0.49$). These low performance reflect a strong model bias toward positive classifications, severely hindering the detection of rare incorrect facts. In comparison, DBpedia yields intermediate results, with balanced accuracy values ranging from $0.54$ to $0.67$ across models.

Among the datasets, FactBench appears to be the most favorable for internal knowledge evaluation, as most models maintain a reasonable balance between \textcolor{black}{$F1(T)$ and $F1(F)$. On the other hand, YAGO proves to be the most challenging due to its large nomber of correct facts. While models achieve high $F1(T)$ scores (up to $0.92$), the $F1(F)$ scores are negligible ($0.01$ to $0.03$). This drastic discrepancy indicates a strong model bias toward positive classifications, which hinders the detection of rare incorrect facts in highly imbalanced contexts. In comparison, DBpedia yields intermediate results; most models achieve respectable $F1(T)$ scores $[0.53, 0.89]$, yet they struggle to reliably identify incorrect information, with $F1(F)$ values generally remaining below $0.40$.}

Notably, the few-shot setup (GIV-F) consistently outperforms both DKA and GIV-Z settings. For instance, \textcolor{black}{on FactBench,} Mistral improves from \textcolor{black}{$0.68$} (DKA) to \textcolor{black}{$0.81$} (GIV-Z), \textcolor{black}{while its performance on false claims remains stable around $0.73$.}. These gains are particularly pronounced for mid-tier models, which benefit more from structured prompting and exemplar-based guidance. By contrast, already well-performing models such as Gemma2 show relatively smaller performance gains.

\stitle{\colorbox{black!10}{Finding 1:}}
Open-source models, such as Gemma2 or \textcolor{black}{Mistral}, outperform commercial alternatives like GPT-4o mini when relying exclusively on internal knowledge. Moreover, few-shot prompting consistently enhances performance, although the degree of improvement is influenced by dataset characteristics such as class balance and label distribution.

%\stitle{\colorbox{black!10}{Finding 1:}}
%Open-source models, particularly \textit{Gemma2}, demonstrate superior performance compared to commercial alternatives when relying solely on internal knowledge. Additionally, we find that few-shot prompting consistently provides improvements, but its effectiveness varies significantly based on the dataset's characteristics.

\paragraph{\textbf{RQ2.}}%\label{subsec:rq2}
We evaluate the performance of the \ac{RAG} methodology across all models and datasets, and then compare it against the internal knowledge-based approaches in Table~\ref{tab:evaluation_results-partition}.

Overall, RAG achieves the highest performance across \textcolor{black}{nearly} all experimental settings. In particular, for the FactBench dataset, RAG delivers substantial improvements: for example, \textcolor{black}{Qwen2.5} achieves a \textcolor{black}{$F1(T)$ of 0.89}, compared to $0.55$ in the DKA setting. This trend holds across evaluated models, including GPT-4o mini, which shows a marked increase in performance -- rising \textcolor{black}{more than 25\% in both $F1$ scores} -- when external evidence is incorporated.

However, the impact of RAG varies significantly across datasets. FactBench and \textcolor{black}{YAGO} show the greatest absolute gains, likely due to \textcolor{black}{their} broader diversity of factual content. In contrast, \textcolor{black}{Dbpedia} exhibits minimal improvements or even slight performance degradation in some cases. This may be attributed to \textcolor{black}{schema diversity, which can complicate the retrieval process and diminish the relevance of the extracted evidence.} 
% high prevalence of correct facts in the dataset and their likely presence in the training data of the models, reducing the added value of external retrieval. DBpedia shows moderate gains (typically 3–4\%), suggesting that schema diversity may influence the relevance and utility of retrieved content.

\stitle{\colorbox{black!10}{Finding 2:}} 
Incorporating external evidence via RAG represents a promising path to high-accuracy fact validation. However, its effectiveness is dependent on dataset characteristics.

\begin{table}[t!]
    \small
    \caption{Model \textcolor{black}{alignment} analysis across fact validation methodologies and datasets. \textcolor{black}{Consensus Alignment ($\text{CA}_M$)} measure the percentage agreement between \ac{LLM} predictions and majority vote decisions, with \most{highest} and \least{lowest} performing models highlighted for each method-dataset combination. Tie percentages indicate the frequency of split decisions requiring arbitration.}
    \resizebox{0.99\linewidth}{!}{\begin{tabular}{ll | c | cccc}
        \toprule
        {Dataset} & 
        {Method} & 
        {Ties} &
        
        Gemma2 & Qwen2.5 & Llama3.1 & Mistral \\
        \midrule
        \multirow{4}{*}{FactBench} 
        & DKA & 16\% &
        0.919 & \least{0.861} & 0.906 & \most{0.938} \\
        & GIV-Z & 21\% &
        0.914 & 0.893 & \most{0.913} & \least{0.814} \\
        & GIV-F & 14\% &
        \most{0.937} & \least{0.861} & 0.901 & 0.909 \\
        & RAG & 6\% &
        0.968 & \most{0.970} & \least{0.897} & 0.960 \\ \hline
        \multirow{4}{*}{YAGO} 
        & DKA & 19\% &
        0.798 & \least{0.797} & 0.916 & \most{0.920} \\
        & GIV-Z & 26\% &
        \least{0.790} & 0.872 & 0.859 & \most{0.886} \\
        & GIV-F & 16\% &
        0.934 & \least{0.771} & 0.901 & \most{0.944} \\
        & RAG & 6\% &
        0.968 & 0.969 & \least{0.916} & \most{0.974} \\ \hline
        \multirow{4}{*}{DBpedia} 
        & DKA & 17\% &
        \most{0.937} & \least{0.772} & 0.891 & 0.920 \\
        & GIV-Z & 24\% &
        \most{0.948} & 0.875 & \least{0.765} & 0.758 \\
        & GIV-F & 17\% &
        \most{0.960} & 0.879 & \least{0.779} & 0.876 \\
        & RAG & 9\% &
        0.953 & \most{0.961} & \least{0.848} & 0.945 \\ 
        \bottomrule
    \end{tabular}}
    \label{tab:evaluation_results-partition11}
\end{table}

\begin{table}[t!]
\small
\caption{\textcolor{black}{Performance evaluation of fact verification systems. The assessment covers multi-model consensus. In each column, the best-performing method is highlighted in bold, and the second-best method is underlined.}}
        \resizebox{0.99\linewidth}{!}{\begin{tabular}{ll | cc | cc || cc}
            \toprule
            \multirow{2}{*}{Dataset} & 
            \multirow{2}{*}{Method} & 
            \multicolumn{2}{c|}{\makecell{agg-cons up \\ \tiny (Refer to Tab.\ref{tab:evaluation_results-partition11})} } & 
            \multicolumn{2}{c||}{\makecell{agg-cons down \\ \tiny (Refer to Tab.\ref{tab:evaluation_results-partition11})} } & 
            \multicolumn{2}{c}{\makecell{agg-\\GPT-4o mini}} \\
            \cmidrule(lr){3-4} 
            \cmidrule(lr){5-6} 
            \cmidrule(lr){7-8} 
            & & 
            $F1(T)$ & $F1(F)$ &
            $F1(T)$ & $F1(F)$ &
            $F1(T)$ & $F1(F)$   \\
            \midrule
            % \rowcolor{gry}
            \multirow{4}{*}{FactBench}  & DKA   & 0.68 & 0.75 & 0.69 & 0.75 & 0.69 & 0.75 \\
                                        & GIV-Z & 0.74 & 0.76 & 0.64 & 0.74 & 0.63 & 0.74 \\
                                        & GIV-F & \underline{0.82} & \underline{0.78} & \underline{0.81} & \underline{0.79} & \underline{0.80} & \underline{0.79} \\
                                        & RAG   & \textbf{0.91} & \textbf{0.89} & \textbf{0.91} & \textbf{0.89} & \textbf{0.91} & \textbf{0.89} \\\hline
            \rowcolor{gry}
            \multicolumn{2}{c|}{Mean}   & 0.79 & 0.80 & 0.76 & 0.79 & 0.76 & 0.79 \\ \hline
            \multirow{4}{*}{YAGO}       & DKA & 0.59 & \textbf{0.02} & 0.63 & \textbf{0.02} & 0.61 & \textbf{0.02} \\
                                        & GIV-Z & 0.63 & \textbf{0.02} & 0.73 & \textbf{0.02} & 0.65 & \textbf{0.02} \\
                                         & GIV-F & \underline{0.84} & \textbf{0.02} & \underline{0.84} & \textbf{0.02} & \underline{0.84} & \textbf{0.02} \\
                                         & RAG & \textbf{0.93} & \textbf{0.02} & \textbf{0.94} & \textbf{0.02} & \textbf{0.93} & \textbf{0.02} \\\hline
            \rowcolor{gry}
            \multicolumn{2}{c|}{Mean} & 0.75 & 0.02 & 0.78 & 0.02 & 0.76 & 0.02 \\ \hline
            \multirow{4}{*}{DBpedia} & DKA & \underline{0.84} & 0.37 & 0.80 & \underline{0.37} & 0.78 & 0.37 \\
             & GIV-Z & 0.77 & 0.38 & 0.73 & 0.36 & 0.71 & 0.36 \\
             & GIV-F & \textbf{0.85} & \textbf{0.40} & \textbf{0.86} & \textbf{0.39} & \textbf{0.81} & \underline{0.38} \\
             & RAG & 0.80 & \underline{0.39} & \underline{0.81} & \textbf{0.39} & \underline{0.80} & \textbf{0.39} \\\hline
             \rowcolor{gry}
             \multicolumn{2}{c|}{Mean} & 0.81 & 0.39 & 0.80 & 0.38 & 0.77 & 0.38 \\
                        \bottomrule
        \end{tabular}}
    \label{tab:evaluation_results-partition-2}
\end{table}

\paragraph{\textbf{RQ3.}}%\label{subsec:rq3}
We investigate the effectiveness of multi-model consensus strategies, applying majority voting across our four open-source models. In cases of ties, we introduce a tie-breaking mechanism using either higher-parameter variants or a commercial model (GPT-4o mini). Table~\ref{tab:evaluation_results-partition-2} summarizes the results.

%We implement majority voting strategies using our four open-source models, with tie-breaking performed by either higher-parameter variants or commercial models. The results are presented in Table~\ref{tab:evaluation_results-partition-2}.

Multi-model consensus provides more reliable performance across internal knowledge settings (DKA, GIV-Z, and GIV-F), although it does not consistently outperform all individual models. In many cases, it stabilizes performance across varying conditions rather than providing top results. Interestingly, the choice of tie-breaking model has minimal influence on final performance. Whether we use the most consistent model (agg-cons-up), the least consistent model (agg-cons-down), or GPT-4o mini, the resulting scores remain nearly identical across all datasets and methods. This suggests that the majority vote mechanism effectively captures the most reliable signal, and the specific choice of arbitrator is less impactful than having a consistent tie-resolution strategy in place.

%Multi-model consensus consistently outperforms individual models, with average balanced accuracy improvements of $\sim$4\% in knowledge-constrained scenarios (DKA, GIV-Z, GIV-F). Remarkably, the choice of tie-breaking model has minimal impact on final performance: whether using the most consistent model, the least consistent model, or the \textit{GPT-4o mini} as the arbitrator. The results remain nearly identical across all methods and datasets.  This finding indicates that the majority vote captures the most available signal, and the specific choice of arbitrator is less important than having a systematic tie-breaking mechanism.

Our consistency analysis, shown in Table~\ref{tab:evaluation_results-partition11}, further reveals that agreement among models increases with methodological complexity. For instance, RAG results in lower tie rates -- ranging from 6\% to 9\% -- compared to 21\% to 26\% in GIV-Z. This reinforces the notion that external evidence not only improves individual model performance but also enhances cross-model alignment. However, this increased agreement may also reflect a stronger influence of shared contextual evidence, potentially reducing reliance on internal knowledge and thereby introducing uniformity at the cost of model individuality or specificity.

%The consistency analysis, as represented in Table~\ref{tab:evaluation_results-partition11}, demonstrates a significant increase in agreement among models in relation to the complexity of the methodology used. Notably, RAG exhibits lower tie percentages, ranging from 6\% to 9\%, in contrast to GIV-Z's tie percentages of 21\% to 26\%. This finding suggests that incorporating external evidence enhances individual model performance and fosters consensus across models, underscoring that high-quality evidence leads to more reliable and harmonized decisions. We must consider that uniformity across models might obscure an elevated bias towards the contextual evidence supplied. This uniformity might arise from a reduced utilization of internal knowledge, thereby diminishing a model's specificity.

\begin{figure*}[!t]
    \centering
    \includegraphics[width=0.49\linewidth,interpolate=False]{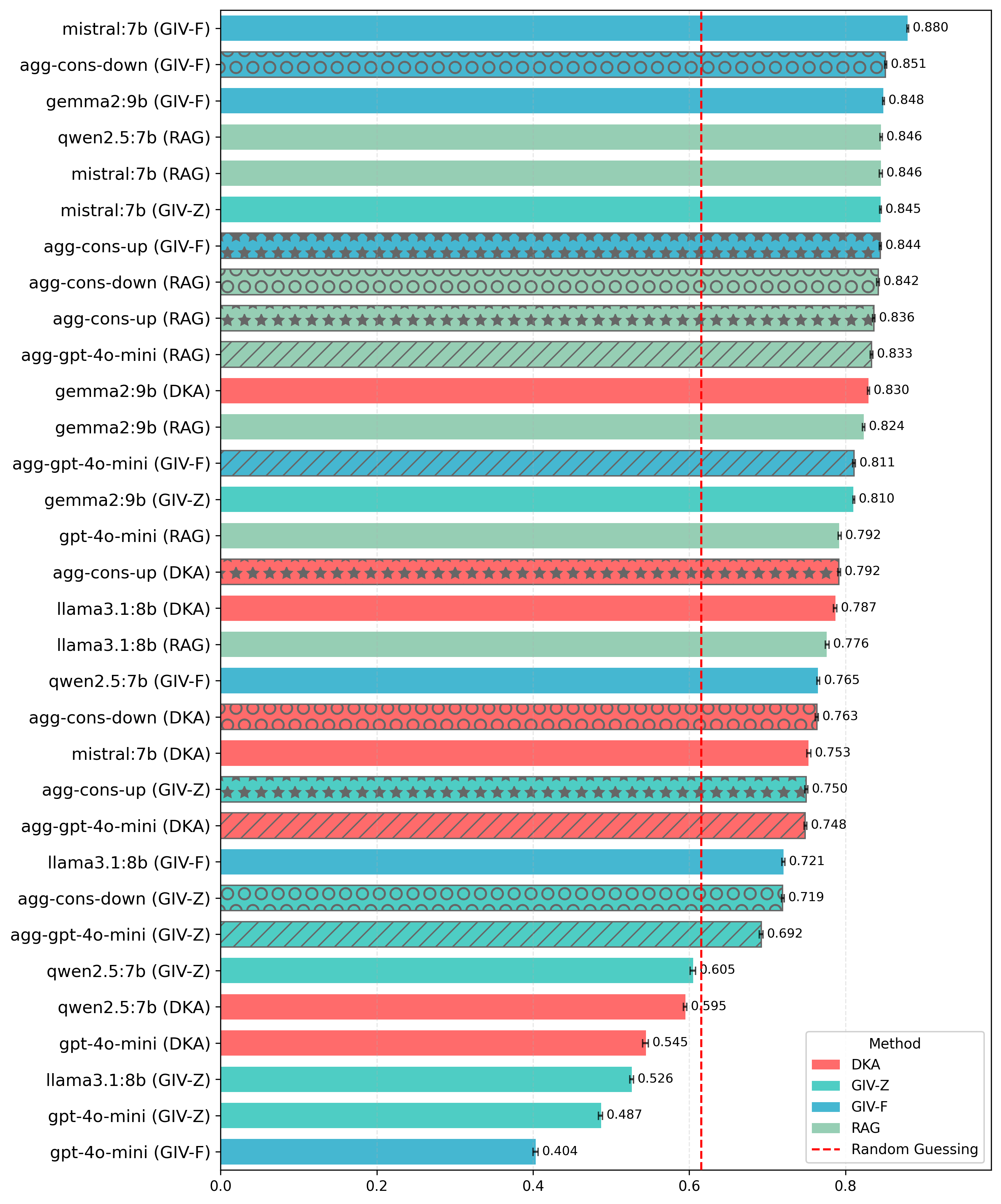}
    \hfill
    \includegraphics[width=0.49\linewidth,interpolate=False]{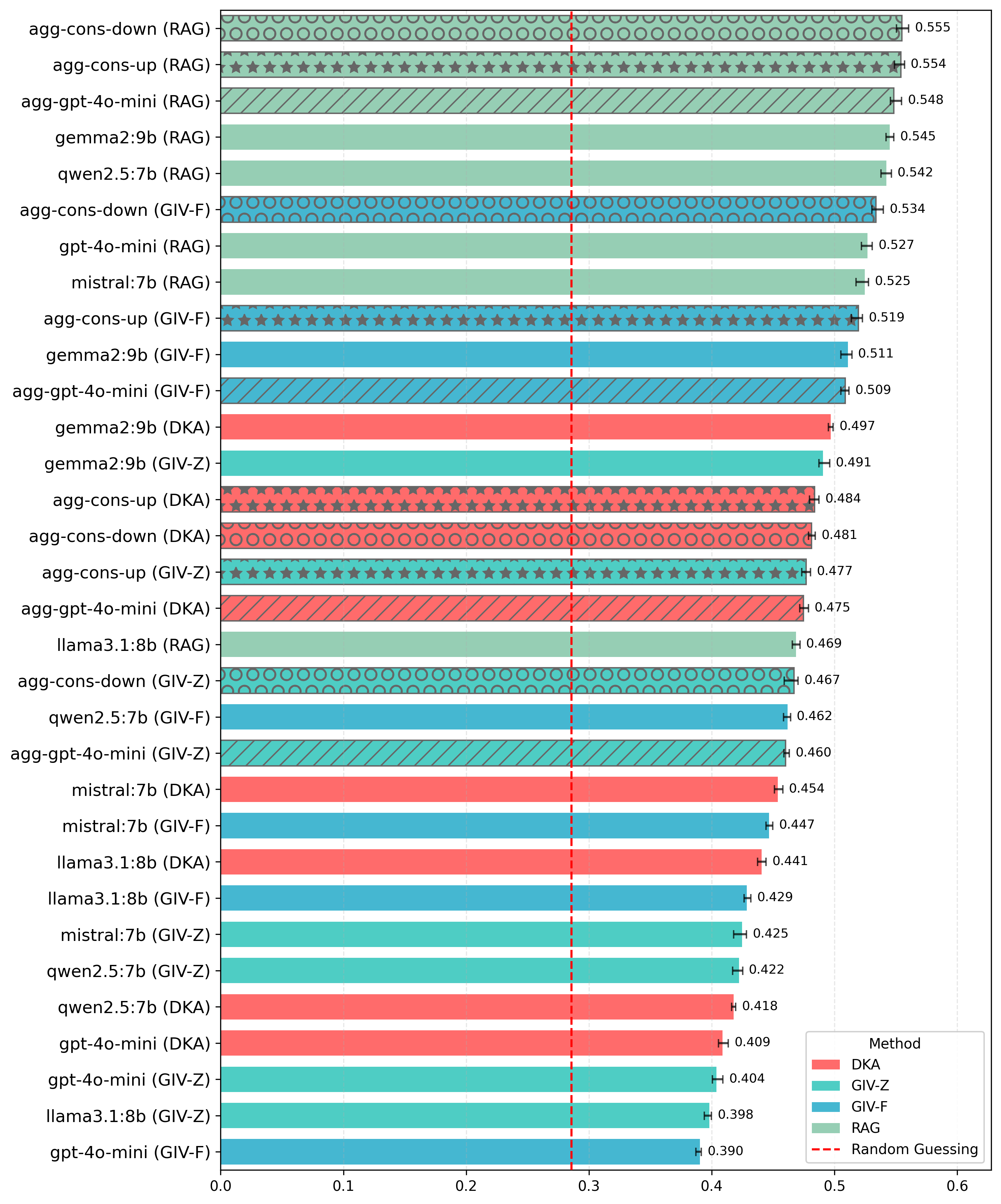}
    \caption{\textcolor{black}{$F1$ scores for \approach benchmark. The left plot displays $F1(T)$ scores, and the right plot displays $F1(F)$ scores. Multi-model consensus results are shown with hatching, and the red dotted line indicates the guess rate.}}
    \label{fig:FactCheckSystehhm3}
\end{figure*}

\stitle{\colorbox{black!10}{Finding 3:}} 
Multi-model consensus offers a simple yet robust mechanism to %mitigating individual model errors
stabilize fact validation performance. While it does not always outperform individual models, it mitigates the impact of weaker ones. The specific choice of arbitrator has a limited impact. Moreover, external evidence promotes greater model alignment, though care must be taken to avoid overfitting to contextual bias.

%The use of majority voting brings robust performance, and while tie-breaking is necessary, the specific choice of arbitrator has limited impact. Moreover, external evidence promotes greater model alignment, though care must be taken to avoid overfitting to contextual bias.

%Multi-model consensus provides robust error mitigation with minimal additional complexity. The majority vote mechanism is more critical than the specific tie-breaking strategy, and external evidence improves both accuracy and model agreement.

\paragraph{\textbf{Computational Efficiency.}}
Beyond accuracy metrics, we evaluate the computational efficiency of different approaches.
Table \ref{tab:execution_times_only} reports execution times (\(\bar{\theta}\), in seconds) for fact validation using the four open-source LLMs across the three reference datasets. Within each dataset, DKA yields the lowest execution times, ranging from $0.21$ to $0.30$ seconds on FactBench, from $0.19$ to $0.31$ seconds on YAGO, and from $0.24$ to $0.37$ seconds on DBpedia. GIV-Z shows an increase over DKA, with approximately double the execution time on FactBench and YAGO, such as an increase from $0.18$ to $0.40$ seconds on Qwen2.5 for FactBench. GIV-F requires more time than GIV-Z, with values reaching up to $0.78$ seconds. RAG results in the highest execution times across all datasets and models, with values including $2.73$ seconds on Llama3.1 for FactBench and over $2.5$ seconds for several models on DBpedia.

\begin{table}[t]
    \small
    \caption{Execution time (\(\bar{\theta}\), in seconds) for fact validation across different methodologies (DKA, GIV-Z, GIV-F, and RAG). The \most{fastest} configuration is highlighted in green, while the \least{slowest} configuration is marked in red.}
    \centering
    \resizebox{0.85\linewidth}{!}{\begin{tabular}{ll | c | c | c | c}
        \toprule
        Dataset & Method & Gemma2 & Qwen2.5 & Llama3.1 & Mistral \\
        \midrule
        \multirow{4}{*}{FactBench} 
        & DKA     & 0.21 & 0.18 & \least{0.30} & \most{0.17} \\
        & GIV-Z   & \least{0.62} & \most{0.40} & 0.50 & 0.45 \\
        & GIV-F   & \least{0.78} & \most{0.51} & 0.67 & 0.65 \\
        & RAG     & 2.27 & 2.39 & \least{2.73} & \most{1.69} \\
        \midrule
        \multirow{4}{*}{YAGO} 
        & DKA     & 0.22 & \most{0.19} & \least{0.31} & \most{0.19} \\
        & GIV-Z   & \least{0.62} & \most{0.41} & 0.45 & 0.47 \\
        & GIV-F   & \least{0.78} & \most{0.54} & 0.69 & 0.67 \\
        & RAG     & 2.10 & 2.39 & \least{2.68} & \most{1.63} \\
        \midrule
        \multirow{4}{*}{DBpedia} 
        & DKA     & 0.35 & 0.25 & \least{0.37} & \most{0.24} \\
        & GIV-Z   & \least{0.70} & \most{0.43} & 0.58 & 0.53 \\
        & GIV-F   & \least{0.89} & \most{0.56} & 0.69 & 0.78 \\
        & RAG     & 2.55 & 2.55 & \least{2.87} & \most{1.77} \\
        \bottomrule
    \end{tabular}}
    \label{tab:execution_times_only}
\end{table}

The comparison within each dataset indicates that, as expected, RAG incurs the greatest computational cost, often exceeding DKA by a factor of six or more. The increase in execution time follows the progression from DKA to GIV-Z to GIV-F to RAG in all configurations. This pattern suggests a direct relationship between the methodological complexity of the verification strategy and its computational cost.

On a different note, multi-model consensus can be parallelized, meaning that inference latency is bounded by the slowest model rather than the sum of all models. In practice, if models exhibit varying response times (e.g., $0.3$–$0.5$ seconds), consensus inference requires waiting for the slowest response, resulting in slightly higher latency compared to selecting only the fastest model. Tie-breaking further adds inference overhead, as it requires an additional model query. Moreover, the coordination and resource allocation across multiple models introduce minor but non-negligible computational overhead. Despite this, consensus brings benefits: %while accuracy gains may range from 2–4\%, 
the trustworthiness of the predictions increases due to the aggregation of diverse model perspectives.

%However, this approach typically costs 1-1.5$\times$ the time of a single model inference because the consensus latency is bound by the slowest model in the ensemble, while individual model selection would allow choosing faster alternatives. For instance, if individual models perform varying response times (e.g., 0.3s to 0.5s), the parallel consensus approach requires waiting for the slowest model to complete and results in longer processing times compared to deploying only the fastest individual model. Furthermore, tie-breaking scenarios require additional model inference, and the allocation of computational resources across multiple models introduces marginal overhead. This issue contributes to the observation of a higher latency factor. The accuracy variations are between 2-4\%, but the important point is that the answers can be more trusted because they are a consensus of the opinions of multiple models.

%The analysis reveals distinct efficiency profiles suitable for different applications: DKA is best for high-throughput scenarios with moderate accuracy requirements, GIV methods offer balanced trade-offs, and RAG is most ideal for tasks where accuracy is paramount.

% \highlight{black!10}{\textbf{Finding 4:}}

\begin{figure*}[ht]
    \centering
    \includegraphics[width=\linewidth]{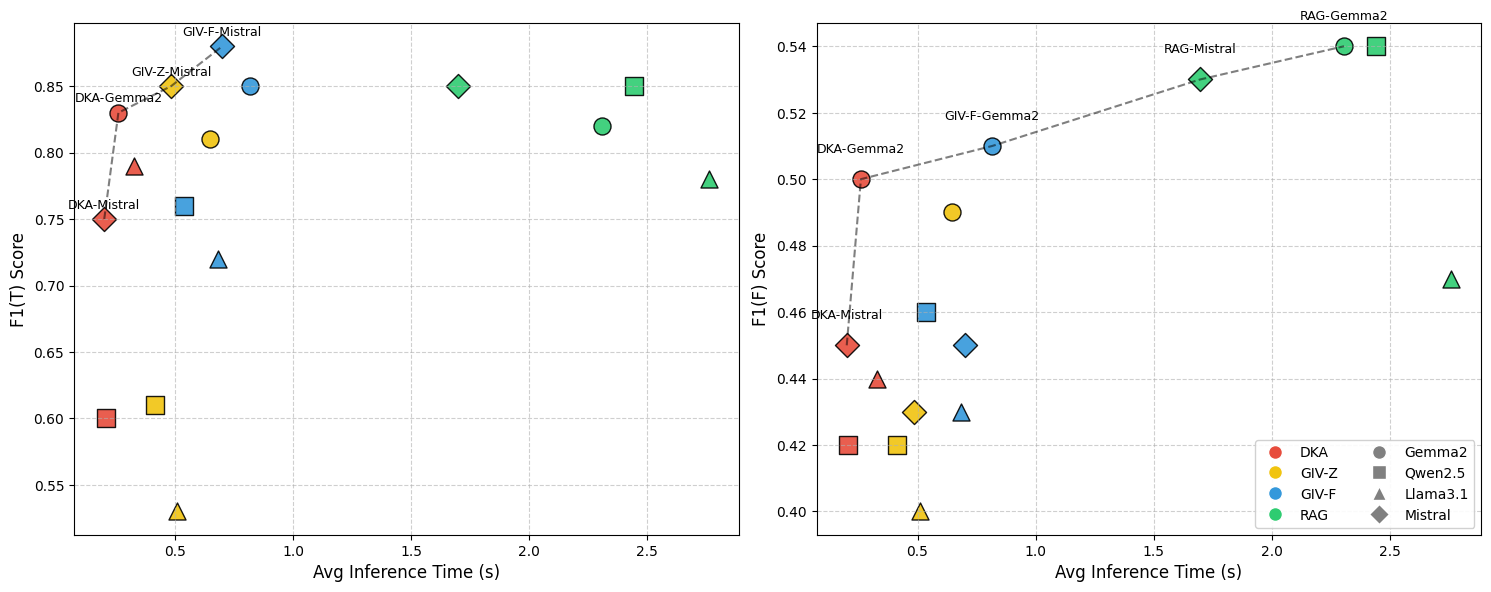}%
    \caption{\textcolor{black}{Trade-off analysis between computational cost ($\bar{\theta}$) and verification performance ($F1(F)$ and $F1(T)$). The dashed line represents the Pareto frontier, highlighting configurations that achieve optimal efficiency (highest accuracy for a given time budget).}}
    \label{fig:trade_off}
\end{figure*}

\textcolor{black}{To characterize the balance between predictive accuracy and computational expense, we examined the Pareto efficiency of our methods across the different models (Figure~\ref{fig:trade_off}). This analysis reveals a clear separation in the utility of each strategy: RAG-based techniques generally cluster in the upper-right quadrant, especially with respect to the $F1(F)$ metric, indicating that their increased latency ($\approx$1.6s–2.9s) is exchanged for enhanced detection of false claims. Conversely, DKA setups dominate the high-speed regime, delivering sub-second inference times ($<$0.3 s) that are appropriate for latency-sensitive use cases, albeit with lower sensitivity. The Pareto frontier indicates that mid-range approaches such as GIV-F (particularly when paired with Gemma2 and Mistral) strike an attractive trade-off, attaining competitive accuracy -- at times even exceeding RAG on the $F1(T)$ metric -- while incurring substantially less computational cost than full retrieval-based systems.}

\stitle{\colorbox{black!10}{Finding 4:}} Computational efficiency varies widely across methods. On the one hand, RAG requires up to 10$\times$ more processing time compared to internal knowledge approaches. On the other hand, consensus strategies can be parallelized to ensure only modest latency increases with respect to internal knowledge methods.

\paragraph{\textbf{Cross-Dataset Generalization and Stability.}}
\textcolor{black}{To assess the generalization capabilities and stability of \ac{LLM}-based fact validation, we analyze the performance across different methods and aggregation strategies, which are visualized in the bar charts (Figure~\ref{fig:FactCheckSystehhm3}). The plots display the $F1$ scores for the True class (left chart) and False class (right chart) ranked by performance. The red dashed line represents the Random Guessing baseline, which sits at approximately $0.62$ for $F1(T)$ and $0.29$ for $F1(F)$, and this reflects the underlying class distribution challenges in the dataset.}

\textcolor{black}{\ac{RAG} demonstrates the most consistent robustness. In the $F1(F)$ chart, which typically represents the harder task of identifying incorrect facts, RAG-based methods and their aggregations dominate the top rankings. On the other hand, GIV-F (blue bars) exhibits high variance. Although Mistral (GIV-F) achieves the absolute highest peak in the $F1(T)$ chart ($0.88$), other models using the same strategy, such as gpt-4o-mini, perform drastically lower at $0.40$. This result falls significantly below the random guessing baseline and suggests that while GIV-F can prompt high recall for true facts in specific models, it lacks the stability of \ac{RAG}. The DKA (red bars) methodology generally occupies the middle-to-lower tier, particularly in the $F1(F)$ analysis, which indicates that reliance on internal parametric knowledge alone is often insufficient for distinguishing false claims. Finally, the aggregation methods denoted as ``agg-cons-$\ast$'' consistently appear in the upper echelons of both charts. This confirms that ensemble reasoning, specifically majority voting strategies, effectively mitigates the volatility of individual models and smoothes out the noise observed in strategies like GIV-Z and GIV-F.}

\stitle{\colorbox{black!10}{Finding 5:}}
\textcolor{black}{\ac{RAG} offers the strongest cross-dataset generalization, consistently outperforming internal knowledge methods in detecting false claims. Some GIV-F models reach top performance on True facts but are highly volatile. Notably, several internal knowledge methods perform below Random Guessing, showing that poor methodology can degrade reasoning to below a coin-flip baseline. Thus, consensus-based aggregation remains essential for stability and reducing model-specific bias.}

\begin{figure*}[htbp]
    \centering
    
    \subfloat[DKA\label{fig:sub1}]{
    \includegraphics[height=5cm]{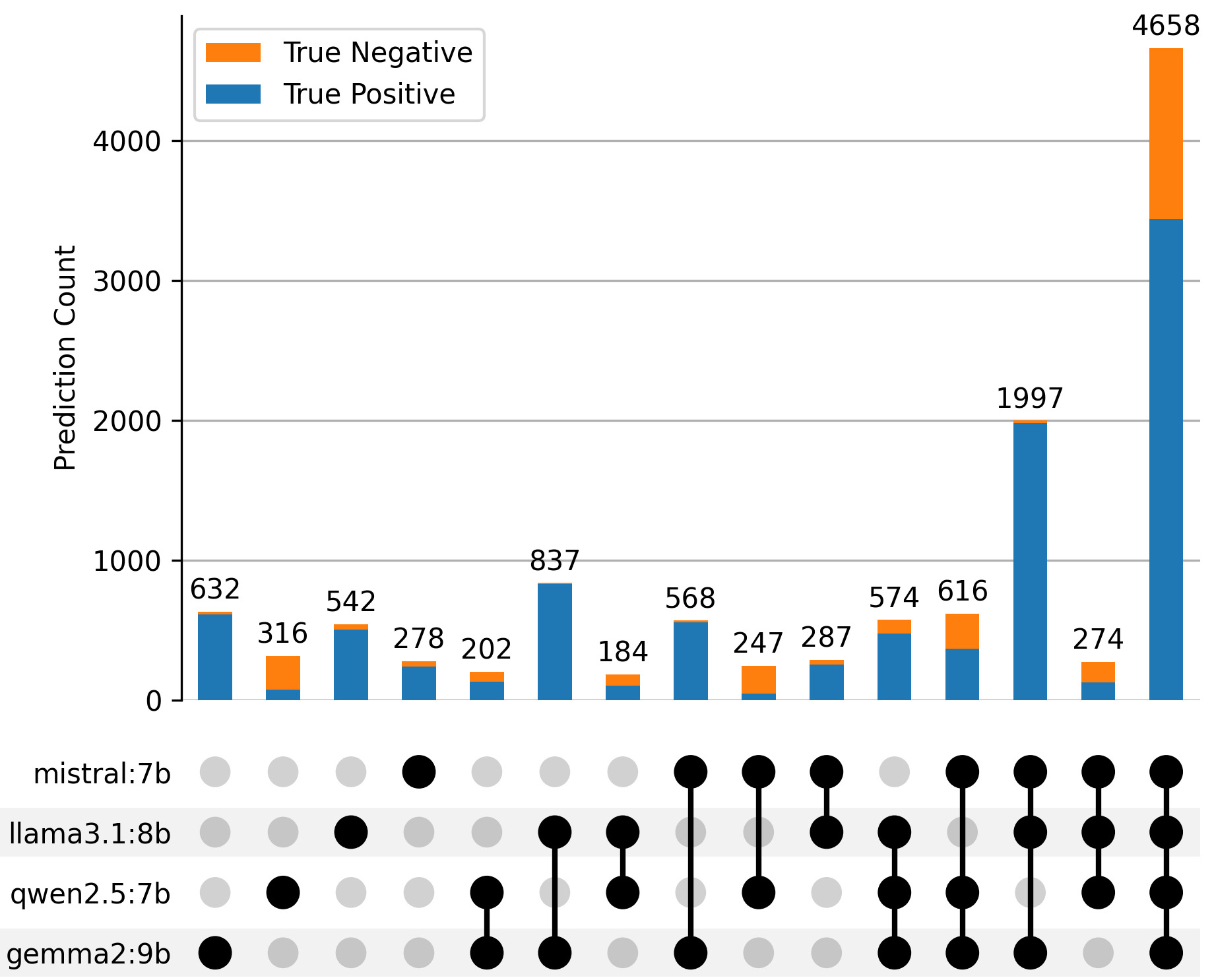}%
    }
    % \hfill
    \subfloat[GIV-Z\label{fig:sub2}]{
    \includegraphics[height=5cm]{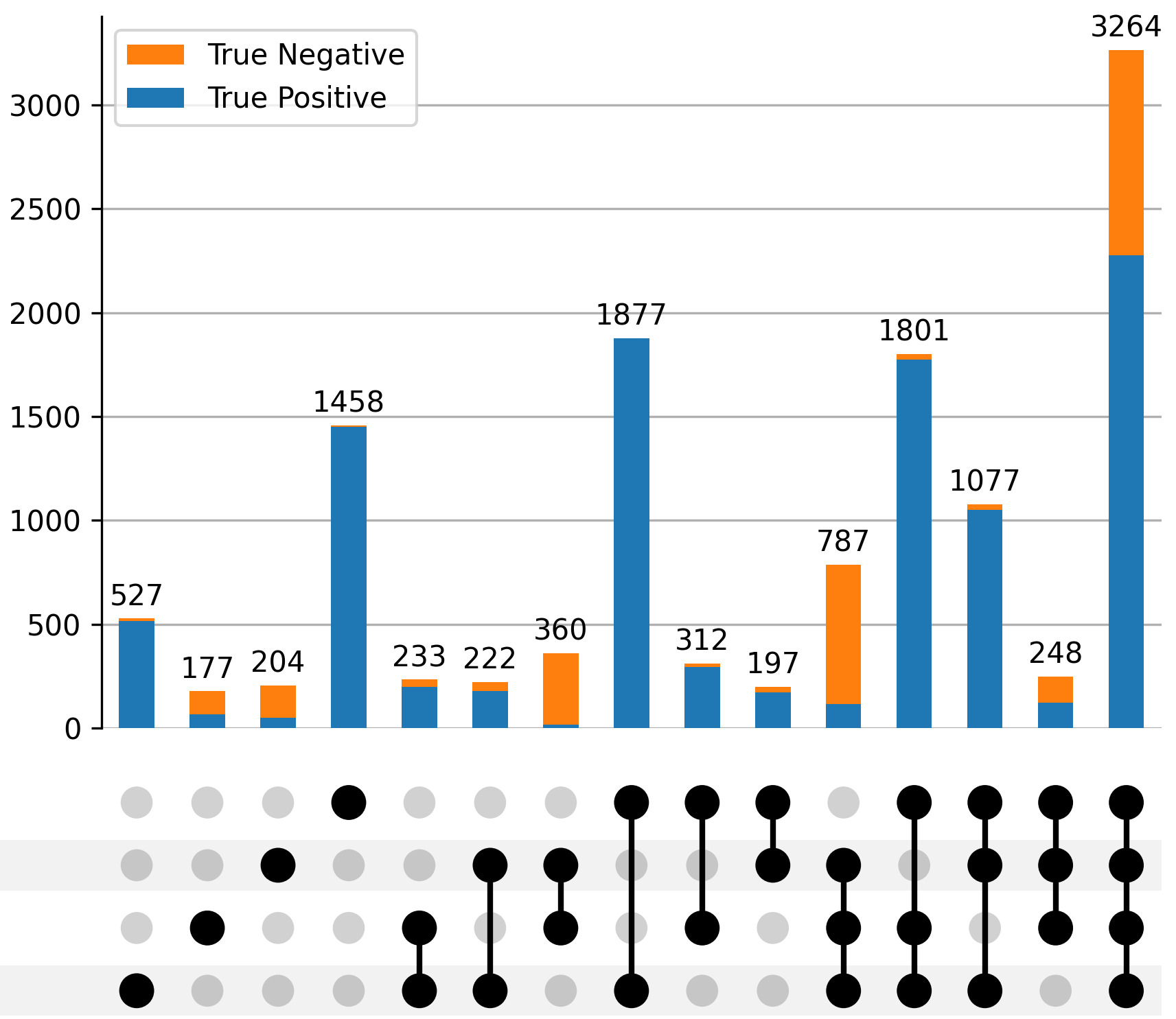}%
    }
    
    % \vspace{0.5cm}
    
    \subfloat[GIV-F\label{fig:sub3}]{
    \includegraphics[height=5cm]{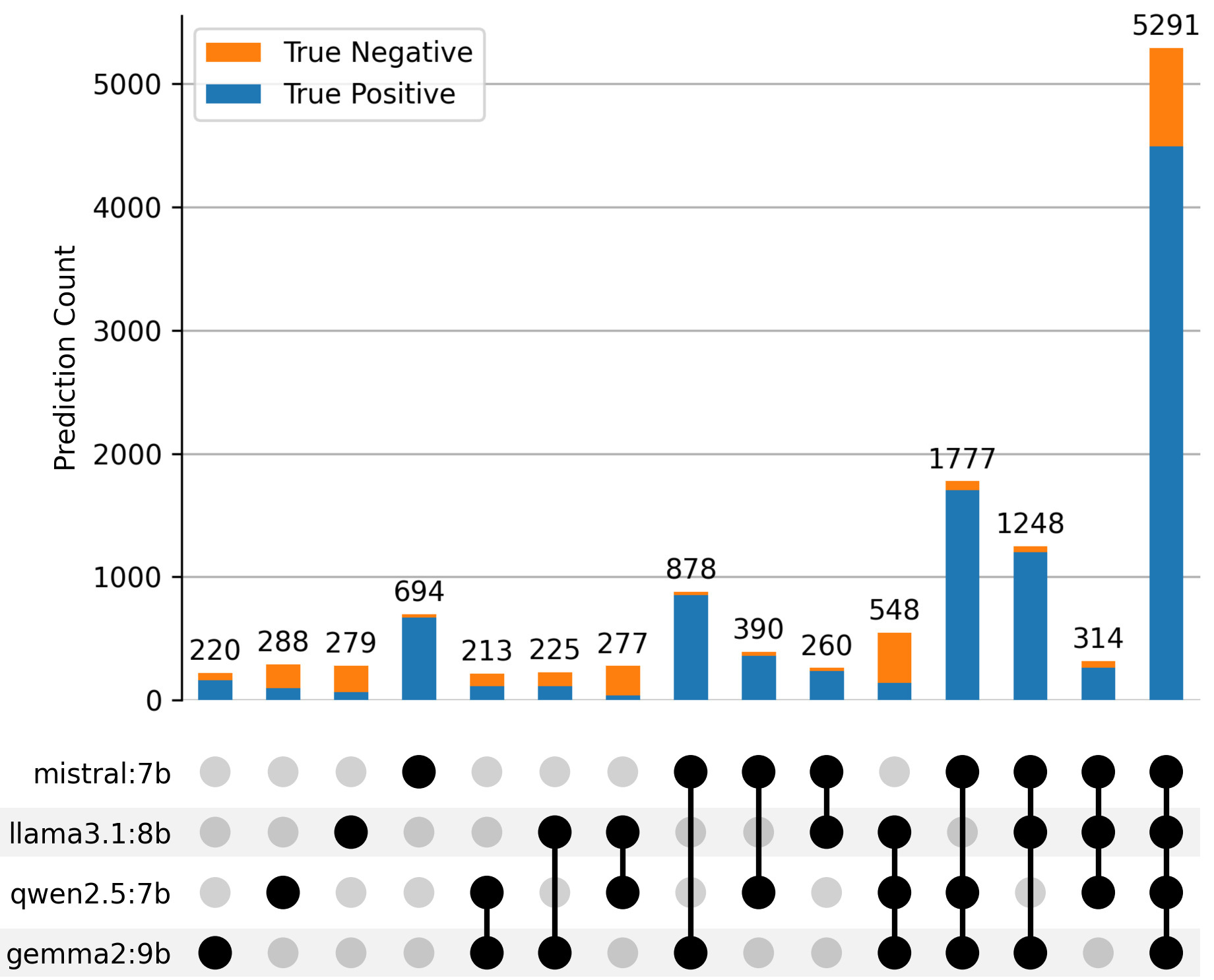}%
    }
    % \hfill
    \subfloat[RAG\label{fig:sub4}]{
    \includegraphics[height=5cm]{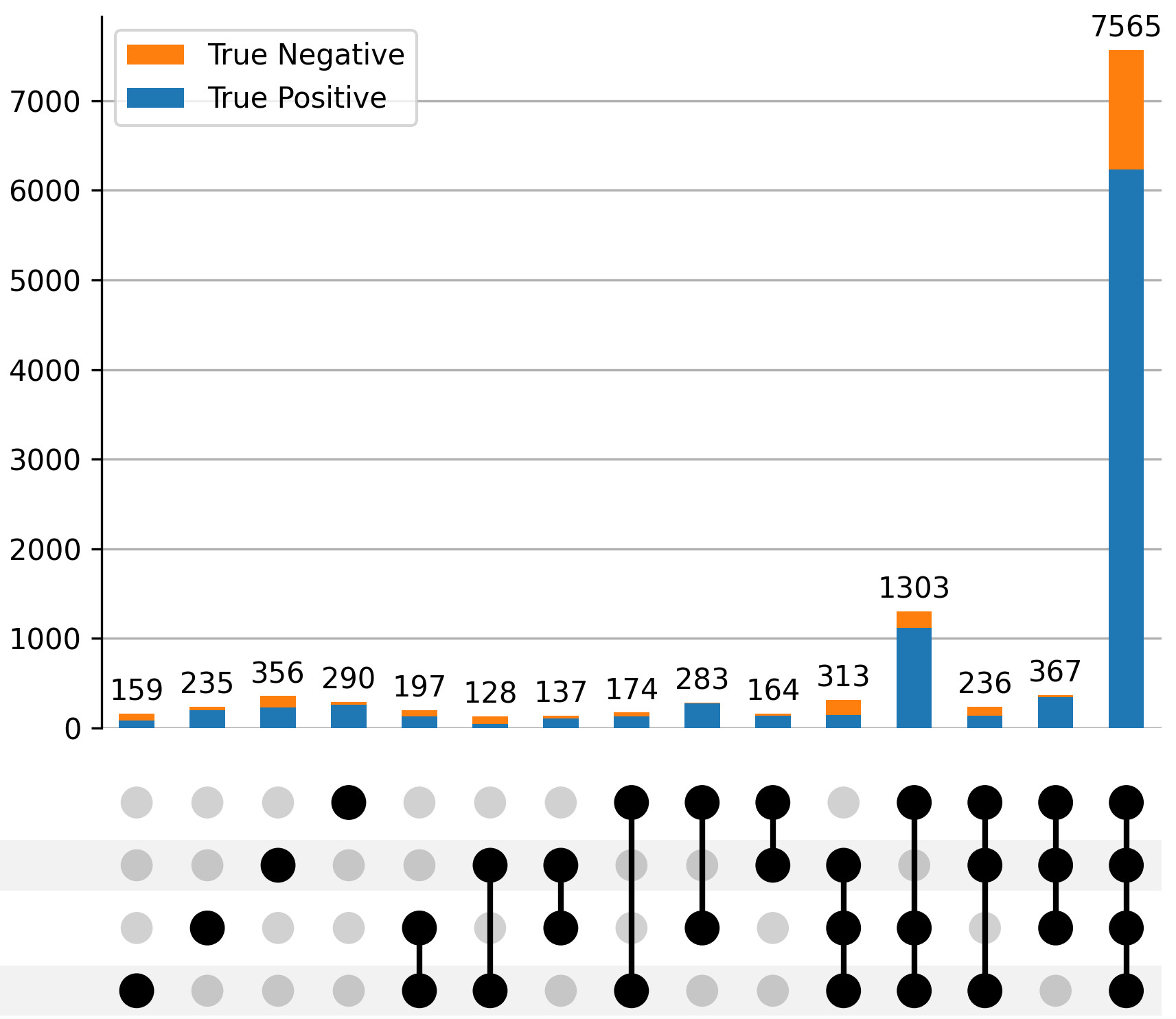}%
    }
    
    \caption{Intersection of correct predictions across models. Bars show the number of correct samples by the specific combination of models indicated by the connected dots below.}
    \label{fig:upset_plots}
\end{figure*}

\section{Qualitative Error Analysis}\label{sec:qualitative_error_analysis}
\textcolor{black}{For our error analysis, we categorize mistakes from open-source models using a semi-automated pipeline combining \ac{LLM}-generated reasoning with contextual document embeddings. We collect logs of incorrect predictions and prompt the same \ac{LLM} to explain each error. Then, we encode these explanations using the \texttt{cde-small-v1} model~\cite{morris2024contextualdocumentembeddings} and cluster them using UMAP for dimensionality reduction followed by HDBSCAN~\cite{CampelloEtAl2013} to find clusters of varying densities. Finally, we assign descriptive labels to each cluster.  
The resulting error categories are:  
Unlabeled (E1): The supplied context is missing the asserted details or mentions of the relevant entities.  
Relationship Errors (E2): The model provides incorrect information about relationships between individuals, such as marital status or religious affiliation.  
Role Attribution Errors (E3): The model wrongly links people to particular roles, locations, or teams.  
Geographic/Nationality Errors (E4): Information about places or national affiliations is inconsistent with the context.  
Genre/Classification Errors (E5): The model miscategorizes movies, genres, or creative works connected to individuals or studios.  
Identifier/Biographical Errors (E6): Identifiers or biographical fact, such as award names, are inaccurate.}

\begin{table}[h]
    \small
    \centering
    \caption{\textcolor{black}{Dataset-wise error clustering based on LLM-generated reasoning.}}
        \resizebox{0.99\linewidth}{!}{\begin{tabular}{ll|cccccc||c}
            \toprule
            \textbf{Dataset} & \textbf{Model} & \textbf{E1} & \textbf{E2} & \textbf{E3} & \textbf{E4} & \textbf{E5} & \textbf{E6} & \textbf{Total}\textsuperscript{*} \\
            \midrule
            \multirow{4}{*}{FactBench} & Gemma2 & 4 & 36 & 45 & 176 & 13 & 1 & 275 \\
            & Qwen2.5 & 33 & 27 & 60 & 194 & 34 & 1 & 349 \\
            & Llama3.1 & 38 & 44 & 73 & 295 & 38 & 3 & 491 \\
            & Mistral & 53 & 27 & 53 & 242 & 40 & 2 & 417 \\ \hline
            \rowcolor{gry}\multicolumn{2}{c|}{Unique. Ratio (\%)} & 0.62 & 0.72 & 0.44 & 0.52 & 0.63 & 0.57 & 0.53 \\\hline
            \multirow{4}{*}{YAGO} & Gemma2 & 6 & 134 & 0 & 14 & 51 & 2 & 207 \\
            & Qwen2.5 & 7 & 109 & 0 & 13 & 63 & 2 & 194 \\
            & Llama3.1 & 8 & 98 & 0 & 19 & 104 & 2 & 231 \\
            & Mistral & 7 & 54 & 0 & 10 & 34 & 3 & 108 \\ \hline
            \rowcolor{gry}\multicolumn{2}{c|}{Unique. Ratio (\%)} & 0.35 & 0.52 & -- & 0.46 & 0.51 & 0.33 & 0.50 \\
            \hline
            \multirow{5}{*}{DBpedia} & Gemma2 & 353 & 22 & 98 & 1729 & 459 & 299 & 2960 \\
            & Qwen2.5 & 339 & 19 & 91 & 1525 & 357 & 237 & 2568 \\
            & Llama3.1 & 382 & 28 & 109 & 2172 & 509 & 318 & 3518 \\
            & Mistral & 325 & 20 & 94 & 1487 & 438 & 241 & 2605 \\ \hline
            \rowcolor{gry}\multicolumn{2}{c|}{Unique. Ratio (\%)} & 0.41 & 0.43 & 0.44 & 0.42 & 0.42 & 0.40 & 0.41 \\
            \bottomrule
        \end{tabular}}
    \vspace{1ex}
    %\parbox{\textwidth}{\footnotesize \textit{Note:} *Some errors aren’t included because we didn’t receive responses.}
    \label{tab:error_results-full-wo-category-all-datasets}
\end{table}

\textcolor{black}{Table~\ref{tab:error_results-full-wo-category-all-datasets} shows the count of each error type on the evaluated datasets. As shown in Table~\ref{tab:error_results-full-wo-category-all-datasets}, E4 errors form the predominant challenge in \approach. In addition, we extended this analysis on the DBpedia dataset using the stratification and topic modeling from Marchesin et al.~\cite{Marchesin_Silvello_Alonso_2024} to understand the impact of fact popularity and domain. The results reveal that error rates decrease in partitions representing common knowledge and domains like ``Education'' and ``News'' yield lower error rates, while ``Architecture'' and ``Transportation'' remain more challenging.} \textcolor{black}{The entire verification process and the error analysis presented here can be interactively interpreted and visualized using our web-based platform available at \url{https://factcheck.dei.unipd.it/}~\cite{10.1145/3726302.3730142}.}

\textcolor{black}{To study how the models complement each other, we examined overlaps in their predictions using UpSet plots~\cite{6876017}. As illustrated in Figure~\ref{fig:upset_plots}, the largest intersection generally corresponds to facts correctly predicted by all four models, indicating that open-source \acp{LLM} share much of their internal knowledge as well as their error profiles. This agreement is most pronounced in the \ac{RAG} setting, where common external evidence steers the models toward the same conclusions, thereby reducing variance. }

\textcolor{black}{GIV-Z, however, departs from this pattern: the ``all-model'' intersection shrinks markedly relative to DKA (from roughly 4{,}600 to about 3{,}200) and is replaced by stronger pairwise overlaps (e.g., between Qwen2.5 and Gemma2). This pattern suggests that zero-shot prompting leads to more heterogeneous reasoning trajectories and greater disagreement among models. In contrast, GIV-F restores stronger consensus, raising the all-model intersection to over 5{,}200, indicating that few-shot demonstrations effectively harmonize model behavior. Overall, the limited true complementarity among models may explain why consensus methods stabilize predictions but rarely outperform the best single model.}

\section{Final Remarks}\label{sec:remarks}
In this work, we introduced \approach, a benchmark for systematically evaluating \acp{LLM} in \ac{KG} fact validation. Our evaluations on three real-world datasets included in \approach -- FactBench, YAGO, and DBpedia -- yielded several key findings. First, open-source \acp{LLM}, such as Gemma2, achieve promising verification performance, with \textcolor{black}{$F1$ scores up to $0.79$ and $0.76$} using internal knowledge alone and exceeding $0.89$ when augmented with RAG. Second, RAG improves performance across most settings, though at a significant computational cost -- being roughly 10$\times$ slower than other methods. Third, multi-model consensus mitigates errors and provides more reliable responses than single-model predictions, in particular when relying on internal knowledge.

At the same time, we also identified several limitations: (1) dataset-specific challenges, such as class imbalance in YAGO and schema diversity in DBpedia; (2) infrastructure constraints, including a 0.08\% retrieval failure rate due to network issues and regional restrictions; and (3) content filtering in hosted deployments, such as blocked factual content on sensitive topics for Azure's GPT-4o-mini.

%\textcolor{black}{Hence, \approach represents an important step forward in understanding and enhancing the factual reasoning capabilities of \acp{LLM}. Unlike prior benchmarks that primarily target unstructured claims or general-domain QA, our benchmark leverages the structured semantics of \acp{KG}, providing a precise and controlled environment for evaluating fact validation. This facilitates reproducible, fine-grained analyses of model behavior across diverse conditions, including the use of internal knowledge, retrieval effectiveness, and multi-model interactions. As such, \approach serves as a robust testbed for researchers developing new prompting strategies, model architectures, or retrieval techniques tailored to fact validation. By releasing our benchmark publicly, we aim to foster transparency, collaboration, and accelerated progress towards trustworthy, scalable KG validation systems.}

\textcolor{black}{Hence, \approach advances the study of \acp{LLM} factual reasoning by leveraging the structured semantics of \acp{KG}, unlike prior benchmarks focused on unstructured claims or general-domain QA. It provides a controlled environment for reproducible, fine-grained analyses of model behavior, including internal knowledge use, retrieval effectiveness, and multi-model interactions. As a robust testbed, \approach supports the development of new prompting strategies, model architectures, and retrieval techniques for fact validation. By releasing it publicly, we aim to promote transparency, collaboration, and faster progress toward trustworthy, scalable \ac{KG} validation systems.}

Looking ahead, our findings suggest several promising research directions. First, fine-tuning or pretraining \acp{LLM} for \ac{KG} fact validation could help mitigate limitations from imbalanced datasets. Second, hybrid retrieval strategies that combine structured KG traversal with unstructured web data may enhance retrieval quality, particularly for datasets like DBpedia. \textcolor{black}{Finally, the benchmark can be extended to support the evaluation of fact-verification systems that also leverage logical rules in the KG, for example by exploiting the ontologies on which the KG is based (e.g., using transitivity, domain/range constraints, and other properties to assess the correctness and reliability of triples).}

\section*{Acknowledgments}
This work is partially supported by the HEREDITARY Project, as part of the European Union's Horizon Europe research and innovation program under grant agreement No. GA 101137074.
The authors thank Andrea Segala for contributing to the experiments on zero-shot and few-shot prompting during his master's thesis.

\section*{Artifacts}
The source code and datasets have been made publicly available at \url{https://github.com/FactCheck-AI/} and \url{https://huggingface.co/FactCheck-AI}.

\newpage

\bibliographystyle{ACM-Reference-Format}
\bibliography{sample-base}

\end{document}